\documentclass[manuscript,nonacm]{acmart}

\setcopyright{none}
\renewcommand\footnotetextcopyrightpermission[1]{}

\usepackage{xcolor}
\usepackage{multirow}
\usepackage{longtable}
\usepackage{array}
\usepackage{graphicx}
\usepackage{float}
\usepackage{svg}
\usepackage{threeparttable}
\usepackage{booktabs}
\usepackage{makecell}
\usepackage{xltabular}
\usepackage{ragged2e}
\usepackage{subcaption}
\usepackage{comment}
\usepackage{pdflscape}

\AtBeginDocument{%
  \providecommand\BibTeX{{%
    \normalfont B\kern-0.5em{\scshape i\kern-0.25em b}\kern-0.8em\TeX}}%
}

\begin{document}

\title{Understanding Parents’ Complex Views of AI for Children’s Pretend Play}

\author{Sungho "Sander" Oh}
\authornote{Corresponding author.}
\email{sungho.oh@drexel.edu; sungho.oh@alumni.upenn.edu}
\affiliation{%
  \institution{Department of Information Science, Drexel University}
  \city{Philadelphia}
  \state{PA}
  \country{USA}
}

\author{Matt Namvarpour}
\email{mn864@drexel.edu}
\affiliation{%
  \institution{Department of Information Science, Drexel University}
  \city{Philadelphia}
  \state{PA}
  \country{USA}
}

\author{Maxi Heitmayer}
\email{heitmayer@rowan.edu}
\affiliation{%
  \institution{Department of Psychology, Rowan University}
  \city{Glassboro}
  \state{NJ}
  \country{USA}
}

\author{Minahil Khalid}
\email{mk4234@drexel.edu}
\affiliation{%
  \institution{Department of Information Science, Drexel University}
  \city{Philadelphia}
  \state{PA}
  \country{USA}
}

\author{Afsaneh Razi}
\email{ar3882@drexel.edu}
\affiliation{%
  \institution{Department of Information Science, Drexel University}
  \city{Philadelphia}
  \state{PA}
  \country{USA}
}

\begin{abstract}
AI could support children’s pretend play, but it could also direct the play on behalf of children. Whether AI should have roles in children’s lives is controversial because its influence on children remains uncertain. We conducted semi-structured interviews with 10 U.S. parents, each with at least one child aged 4—15. During the interview, we described the concept of AI-supported pretend play and provided participants with two boundary-case storyboards. We analyzed the interview data through codebook thematic analysis, using inductive coding and affinity diagramming organized around the research questions, and then used qualitative systems mapping to examine relationships within and across themes. We found that the same characteristics of AI, e.g., ability to assume characters, responsiveness, and adaptability, were seen by parents as potentially useful but also concerning. Parents imagined that AI could make role-based play accessible to all children or help parents participate in family play. However, they opposed the idea of AI for children’s play without a clear understanding of how it works and its long-term influence on their children. Parents worried about children’s loss of imagination and creativity, emotional attachment to AI, reduced human interaction, inappropriate behavior by AI and/or children, and their inability to manage children’s AI use. Parents viewed AI not only as a play tool but also as a social actor and a possible perturbation in the existing family dynamics. The appropriateness of AI and child--AI interactions therefore emerged as a requirement for AI in children’s pretend play, in addition to technical safeguards and parental control. We contribute an integrated account of parents’ interdependent judgments and emphasize the need for longitudinal research with children and their diverse families.
\end{abstract}



\maketitle

\section{Introduction}

Children engage in pretend play when they use their imagination to transform ordinary objects, actions, people, or situations, e.g., by treating a cardboard box as a spaceship or pretending to be a superhero \cite{Weisberg2015PretendPlay}. Through imagined situations and roles, children can express their ideas and emotions and explore social relationships. Pretend play is associated with children’s social competence, whereas its developmental value depends on the quality and context of play rather than simply how often children play \cite{SmitsVanDerNat2024TheValueOf,Lillard2013TheImpactOf}. In this paper, we use \textit{pretend play} as a broad term that includes various types of play overlapping with pretend play, provided that imagination remains the central component of the activity (Figure \ref{fig:1PretendPlay}).

Children's pretend play has been explored in human--computer interaction (HCI) through interactive objects, digitally augmented toys, and physical--digital play technologies \cite{DeValk2013LeavingRoomFor,Hong2019InvestigatingTheEffect,Torres2021ASystemicReview}. While these technologies can provide materials and feedback that support children’s participation and extend their play, what determines how they contribute to children’s pretend play will depend on how they are designed. Open-ended interactive objects can allow children to create their own play goals, rules, and meanings, whereas directive features may restrict children’s choices and control over the activity \cite{DeValk2013LeavingRoomFor,Torres2021ASystemicReview}. Additionally, overly structured and attention-capturing digital experiences can limit children’s control over play \cite{Colvert2024PlayfulByDesign}, with prolonged screen use displacing important opportunities for social play in children's lives \cite{Putnick2023DisplacementOfPeer}.
 
Conversational and generative AI can respond to children, generate story content, take on characters, and continue an interaction \cite{Fan2024StoryPromptExploringThe,Dangol2026ToysThatListen}. Recent HCI studies suggest that these technologies can support responsive storytelling and child--AI co-creation \cite{ Fan2024StoryPromptExploringThe,Sun2024ExploringParentsNeeds}. In this paper, we use \textit{AI} broadly to refer to technologies as parents might imagine them, ranging from traditional AI (e.g., machine learning), chatbots, and voice-based technologies to conversational and multi-modal generative AI systems. The capabilities of these technologies may support pretend play by allowing AI to enact roles assigned by children, respond to their imagined situations, and offer characters or story ideas when children or parents need support \cite{Fan2024StoryPromptExploringThe,Voysey2026SupportingParentsPlayfulness}. Although AI might have the potential to support more interactive play, particularly when caregivers remain involved \cite{Voysey2026SupportingParentsPlayfulness}, serious concerns about AI's implications for children's safety and long-term development remain unresolved \cite{Yu2025UnderstandingGenerativeAI}.

Introducing AI into children’s pretend play raises important questions not only about AI safety for children \cite{Sun2024ExploringParentsNeeds,Driscoll2026UnderstandingParentsDesires}, but also about how children's interaction with AI can be appropriate. In addition, introducing AI into the play of younger children also means introducing it into the family ecosystem. Research on in-home robots and voice assistants shows that these technologies might become part of family routines, interactions, relationships, and values \cite{Cagiltay2023FamilyTheoriesIn,Huang2026FamilyDynamicsWith}. Thus, children’s use of AI in family settings raises additional safety and privacy concerns, potential disruptions in family relationships, and increasing responsibilities for parents as primary caregivers \cite{McReynolds2017ToysThatListen,Driscoll2026UnderstandingParentsDesires}. 

Recent HCI studies have examined parents’ views of generative AI, including the studies about self-directed learning of children \cite{Xie2026UnderstandingParentsPerspectives}, the moderation of children’s open-ended conversations with generative AI \cite{Driscoll2026UnderstandingParentsDesires}, and the use of generative AI to support parents’ playfulness with their children \cite{Voysey2026SupportingParentsPlayfulness}. These studies have shown that parents’ expectations, concerns, desired safeguards, and willingness to support generative AI for play vary widely across family contexts and even between children within a family. Additionally, it is becoming more prevalent that companies integrate generative AI conversations as part of their products \footnote{https://www.nytimes.com/2025/08/15/arts/ai-toys-curio-grem.html}. Nonetheless, it is not well understood how parents connect AI-supported pretend play with its perceived developmental value, associated risks, parental responsibilities, safety requirements, and design conditions for AI's appropriateness. To learn about how parents perceive and understand the potential roles of AI technology in children’s pretend play, we ask the following research questions:

\begin{description}
    \item[\textbf{RQ1:}] \textit{What are parents' perceptions about the potential roles of AI in children's pretend play?}
    \item [\textbf{RQ2:}] \textit{What types of risks did parents associate with children engaging in pretend play with AI?} 
    \item [\textbf{RQ3:}] \textit{How do parents imagine a safe and appropriate AI design for pretend play?}
\end{description}

To answer these questions, we conducted semi-structured interviews with 10 U.S. parents who had at least one child aged 4--15. We broadly described the concept of AI-supported pretend play, allowing the participants to form and describe their own mental models of the application. We also used storyboards to get responses to two boundary cases. Our analysis began with inductive codebook development following the method of codebook thematic analysis \cite{Roberts2019AttemptingRigourAnd}. We then conducted affinity diagramming of the codes organized around the research questions to refine them into themes \cite{Lucero2015UsingAffinityDiagrams}. Finally, we used integrative qualitative systems mapping to examine relationships within and across themes \cite{Hanger-Kopp2024WhatQualitativeSystems}, which deepened our understanding of the complexity of parents' perceptions of AI in children's pretend play.

This study contributes two distinct insights: a qualitative analysis of parents' perceptions of the concept of AI-supported pretend play as a family game that could support parent--child participation and an appropriateness taxonomy that connects AI’s design and behavior with children’s agency, parental protections, and individual family contexts. 
\section{Background}

\subsection{Pretend Play as an Overarching Term}
\label{sec:pretend-play}

Children's pretend play brings together several overlapping forms of play. \textit{Pretend play} involves non-literal actions, in which children act as if an object, person, or situation is different from reality. For example, a child may pretend to be a doctor or use a block as a telephone \cite{Weisberg2015PretendPlay}. The use of an object or action to represent something else is commonly described as symbolic play. Simple forms of symbolic play emerge around 18 months of age and appear jointly with make-believe play more frequently between ages two and three \cite{Weisberg2015PretendPlay, Ungerer1981DevelopmentalChangesIn}. Based on imaginative components, during the preschool years, pretend play often expands to merge with \textit{role play}, in which children act as characters or take on social roles. When children coordinate roles and imaginary situations with other players, the pretend-play activity becomes social \cite{Bretherton1989PretenseTheForm, Howes1992SequencesInThe}. \textit{Narrative play} connects characters and actions through a storyline that develops in time. Pretend play and narrative play frequently overlap as expressions of children's symbolic imagination, although they are not identical \cite{Nicolopoulou2005PlayAndNarrative}. 

\begin{figure}
    \centering
    \includegraphics[width=0.65\linewidth]{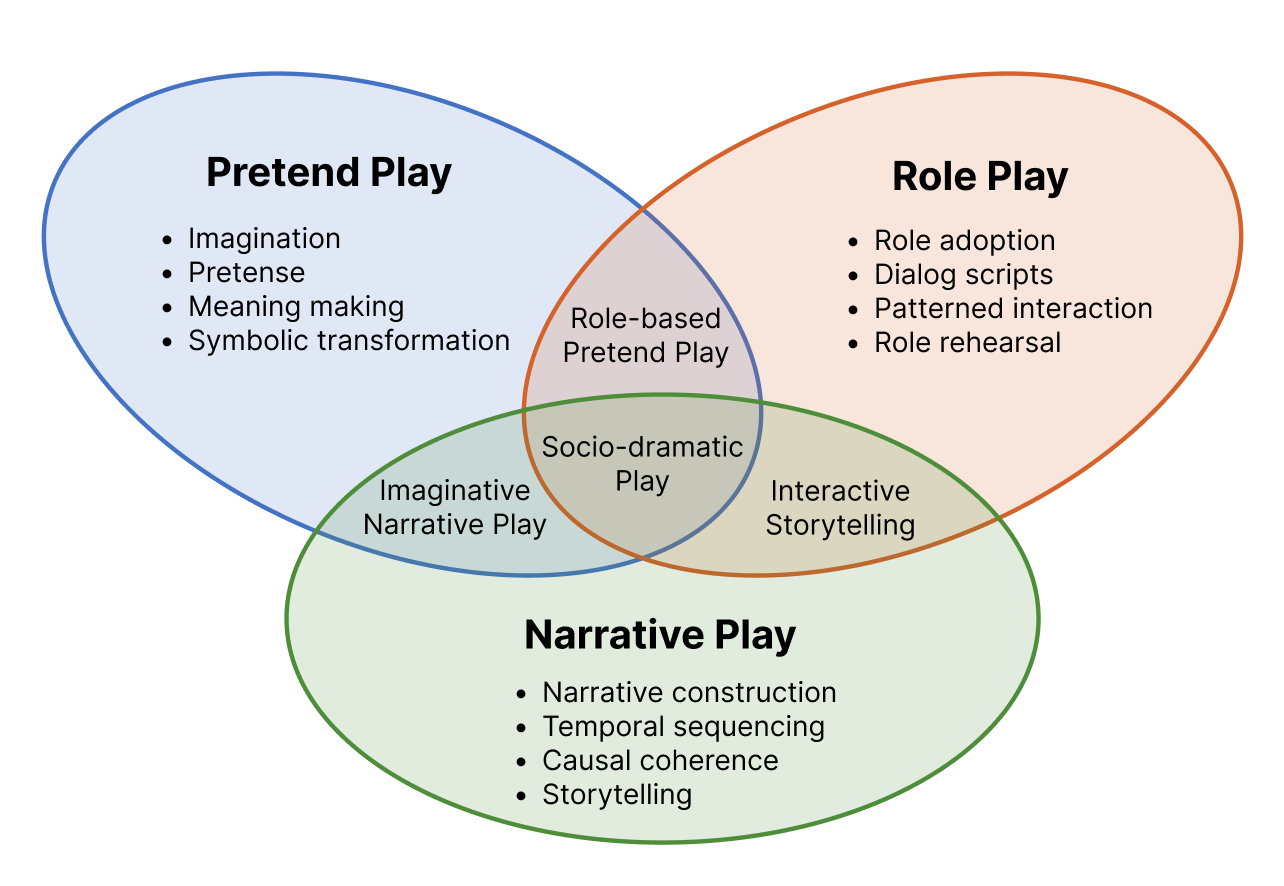}
    \caption{Conceptual Relationships among Pretend Play, Role Play, and Narrative Play}
    \Description{A conceptual diagram showing pretend play as a broad category of imaginative play. Role play and narrative play are shown as related and overlapping forms of pretend play: an activity may involve adopting roles, constructing a narrative, or both.}
    \label{fig:1PretendPlay}
\end{figure}

During children's everyday lives, various forms of play consisting of symbolic actions, make-believe imaginative play, role-based interactions, and narrative interactions tend to occur together rather than as separate activities \cite{Nicolopoulou2005PlayAndNarrative, Bergen2001PretendPlayAnd, Bergen2002TheRoleOf}. For example, children may transform a cardboard box into a clinic, take on the roles of doctors and patients, and create a story about treating an injured animal. However, it is worth noting that role play and narrative play can be different from imaginative play in that the players might passively act on a role, reading off a given script or following another player’s lead without imagining or creating things autonomously. Similarly, a child’s imagination and narrative during play may be interdependent but can also be viewed as distinctly complementary \cite{Nicolopoulou2005PlayAndNarrative, Friedman2007TheConceptualUnderpinnings}. Therefore, this paper uses \textit{pretend play} as an overarching term to include not only imaginative play (e.g., symbolic, make-believe, fantasy play), but also to cover the parts of role play and narrative play where the imagination component enriches children’s play activity. Figure~\ref{fig:1PretendPlay} illustrates these distinctions and the meaning of pretend play used in this paper.

\subsection{Pretend Play: Why do We Care?}

Pretend play is meaningfully associated with several areas of children's development. Creating an imaginary world, children can transform objects, generate ideas, use language, express emotions, solve problems, and explore social situations. Role-based pretend play allows them to consider different perspectives and relationships, while imaginative narrative play helps them connect events, actions, and consequences using narratives. These activities provide opportunities to practice imagination, creativity, communication, social understanding, and emotional control. A meta-analysis of 34 studies found a positive relationship between pretend play and children's social competence \cite{SmitsVanDerNat2024TheValueOf}. Experimental research has also shown that specific forms of dramatic pretend play can support particular skills. For example, a controlled intervention found that dramatic pretend-play games improved emotional self-control among four-year-old children \cite{Goldstein2018DramaticPretendPlay}. Lillard and colleagues questioned whether the existing evidence was sufficient to conclude a causal relationship between pretend play as a single variable and broad developmental outcomes \cite{Lillard2013TheImpactOf}. This caution may not mean that pretend play lacks developmental importance or meaningful associations with children's abilities. Instead, it may suggest that researchers should examine the forms, features, and contexts that produce quality pretend play and support particular developmental opportunities \cite{Jaggy2020TheEmergenceOf,Thompson2019DisengtanglingPretendPlay}.

Researchers distinguish the quality of pretend play, the complexity of what children do, from simply how often or how long they play \cite{SmitsVanDerNat2024TheValueOf}. Quality pretend play integrates imaginative transformation, sustained role enactment, and narrative development, especially when children coordinate these elements with other players \cite{Thompson2019DisengtanglingPretendPlay, Jaggy2020TheEmergenceOf}. Consistent with this view, studies measuring the quality or complexity of pretend play have found stronger relationships with social competence than studies measuring only the amount of pretend play \cite{SmitsVanDerNat2024TheValueOf}. Here, caregivers can facilitate peer play, join the play themselves, or offer play materials and ideas \cite{Weisberg2016GuidedPlayPrinciples, Kalkusch2021PromotingSocialPretend}. A systematic review of guided-play studies has also found that such support can benefit specific areas of children's learning \cite{Skene2022CanGuidanceDuring}. Effective guidance therefore can help children develop their imaginary worlds while allowing them to be the authors and leaders of their play.

Alongside the complexity that enhances the quality of pretend play, children's agency should also matter during the play. To achieve high-quality pretend play, children can use their own imagination and social/narrative capacities to varying degrees; thus, they may have different support needs \cite{Thiemann-Bourque2019ApplicationOfThe, DeKroon2002PartnerInfluencesOn, Currin2021SupportingShyPreschool}. To maintain their agency over play-related cognitive processes and activities, some children may need support in generating or extending imaginary ideas, while others may readily create a play world but find it difficult to join peer play, communicate a role, or coordinate a shared storyline. Shyness or social anxiety, for example, can make joining social pretend play challenging, which can be remedied by supportive technologies, such as a voice agent \cite{Currin2021SupportingShyPreschool}. Language-development delays may affect how easily children explain imagined situations, negotiate roles, or express a developing narrative \cite{Lewis2000RelationshipsBetweenSymbolic}. Similarly, autistic and other neurodivergent children can communicate, imagine, or participate in social play in ways that differ from the expectations of their play partners \cite{Shire2020PeerEngagementIn, Morris2024UnderstandingNeurodiverseSocial}. These differences indicate that children require ways to enter, express, and sustain imaginative, social, and narrative play that meet their individual needs. Thus, accessibility may depend on responsive play partners, suitable forms of communication, adaptable play structures, and sufficient opportunities for children to participate according to their abilities and preferences.


\subsection{How Digital Toys and Screens Shape Children's Play}
\label{sec:toys-screens}

Recent evidence supports caution toward screen technologies designed with children's engagement as a primary goal but without prioritizing children's safety and developmental and learning needs \cite{Mallawaarachchi2024EarlyChildhoodScreen, Hirsh-Pasek2015PuttingEducationIn}. Screen use raises concerns when it becomes prolonged, difficult to stop, or replaces other activities, because excess screen use has been associated with shorter sleep and behavioral concerns such as anxiety, aggression, and inattention \cite{Li2020TheRelationshipsBetween, Eirich2022AssociationOfScreen}. Increased screen use among toddlers has also been associated with less peer play, suggesting that screens may displace opportunities for social interaction \cite{Putnick2023DisplacementOfPeer}. These concerns become greater for designs aimed at holding children's attention only, because apps for young children commonly include rewards that encourage continued play or prompts to make purchases \cite{Radesky2022PrevalenceAndCharacteristics}. 

Toys can also shape how children play, influencing the quality and direction of their play \cite{TrawickSmith2015EffectsOfToys}. Their design features can make it easier for children to initiate play activities; however, highly prescriptive designs may limit children's opportunities to create meanings, roles, and narratives on their own. In contrast, open-ended toys leave more decisions to children; De Valk and colleagues described how open-ended play with interactive objects may help children construct their own rules, goals, and meanings \cite{DeValk2013LeavingRoomFor}. Thus, the design of a toy, physical or digital, can either support or restrict children's agency and imagination.

A systematic review found that physical--digital play technologies could support behaviors such as collaboration, decision-making, problem-solving, physical activity, and self-monitoring; however, these effects depended on the features of the technology such as how directly it controlled children's actions and whether its design aligned with the intended form of play \cite{Torres2021ASystemicReview}. Digital features may similarly support or disrupt pretend play. In an observational study with 32 children aged three to seven, audiovisual features helped children assign new meanings to objects and negotiate shared pretend scenarios but sometimes intensified conflicts between players \cite{Hong2019InvestigatingTheEffect}. Thus, the value of these toys and technologies should depend on whether their design and technological features support children's abilities to create, imagine, and socially interact without taking control of the activity from them. Toys connected to the internet can record and transmit children's speech or other information outside the home, and children may not understand that other people could access what they say to these toys \cite{McReynolds2017ToysThatListen}. Therefore, digital toys should be evaluated to determine whether they preserve child-directed play while avoiding excessive use, manipulative engagement, and unnecessary data collection.

\subsection{AI in Children's Play: Safety Risks Identified in Research on AI for Children}
\label{sec:ai-pretend-play}

AI adds a new layer to digital play. Passive media present children with content, while conventional digital toys usually select from a fixed set of responses. Generative AI systems can respond to children and generate new story content during an interaction \cite{Fan2024StoryPromptExploringThe}. When embedded in toys, these systems may also adopt distinct character personas and use memory across interactions to personalize later responses \cite{Dangol2026ToysThatListen}. A review of 38 HCI studies showed that children, especially older teenagers, already interact with voice-based conversational AI agents for learning, information, entertainment, and social interaction \cite{Garg2022TheLastDecade}.

HCI researchers have begun to explore these technologies in storytelling and creative activities for younger children \cite{Beneteau2020ParentingWithAlexa, Zhang2022_StoryBuddyAHuman, Fan2024StoryPromptExploringThe, Voysey2026SupportingParentsPlayfulness}. For example, StoryBuddy allows parents to create interactive storytelling experiences with AI, supporting joint parent--child reading and more independent use by children \cite{Zhang2022_StoryBuddyAHuman}. StoryPrompt, similarly, allowed elementary-school children to co-create stories and comics with AI \cite{Fan2024StoryPromptExploringThe}. These systems illustrate how AI could provide story ideas, questions, characters, or feedback. Related research has examined how parents and children can use AI together and showed AI's potential to scaffold parental playfulness with their preschool children by providing ideas for parent--child play \cite{Voysey2026SupportingParentsPlayfulness}. These works warrant investigating whether a similar approach could support pretend play and parent--child connectedness.

The same technological capabilities introduce risks beyond prolonged screen time. A recent review identified six broad areas of concern in youth interactions with generative AI: mental well-being, behavioral and social development, toxic content, privacy, bias and discrimination, and misuse \cite{Yu2025UnderstandingGenerativeAI}. Relational design creates an additional concern. Media equation theory proposes that people often apply social expectations to technologies that communicate like people, even when they know that the technologies are not human \cite{Reeves1996TheMediaEquation}. Therefore, a responsive voice, remembered information, and emotional language can encourage children to treat AI as a social actor. Recent studies illustrated why the boundary between a temporary play role and a broader companion role may be difficult to maintain~\cite{Dangol2026ToysThatListen,Namvarpour2026UnderstandingTeenOverreliance}. In a study that employed toys simulating emotions, remembering information, and personalizing later responses, children perceived the toys as social beings even when they did not respond as expected~\cite{Dangol2026ToysThatListen}. Another study showed that the creative or playful use of companion chatbots may develop into strong attachment and overuse for some teens \cite{Namvarpour2026UnderstandingTeenOverreliance}. Although the findings on teenagers cannot be applied directly to younger children, they show why emotional responsiveness and repeated interaction require careful study.

\subsection{Parents' Views of AI in Children's Pretend Play: What Is Known and What Remains Unknown}
\label{sec:why-parents-views}

Introducing AI into children’s play also introduces it into the family ecosystem. Family-centered design proposes that children’s technologies should be considered in relation to family relationships, routines, values, and responsibilities \cite{Cagiltay2023FamilyTheoriesIn}. Families differ in how they use these systems, what roles the systems take in the home, and how family members exercise autonomy \cite{Huang2026FamilyDynamicsWith}. This variability may explain why the same feature is helpful for one family but concerning for another: an AI feature may help one parent join play but create additional work for another. Parents’ views of technology can therefore offer important insights into how AI may fit into, or disrupt, family life. Although parents’ views cannot replace children’s perspectives, they observe and influence their children’s interests, play habits, and development over time. They also make decisions about which technologies enter the home and how their children use them.

Parents do not evaluate children’s technologies in isolation. Instead, they weigh multiple concerns, including perceived utility, safety, privacy, family relationships, the parental work required to manage technology, and personal or family preferences \cite{Cagiltay2023FamilyTheoriesIn,Huang2026FamilyDynamicsWith,Driscoll2026UnderstandingParentsDesires}. Parents have expressed a need for age-sensitive controls over children’s open-ended conversations with generative AI \cite{Driscoll2026UnderstandingParentsDesires}. Managing children’s technology use may also create additional work for parents, who must fit it around other caregiving responsibilities \cite{Hiniker2016ScreenTimeTantrums}. Understanding parents’ views of AI for children therefore requires examining why they may accept, limit, or reject AI rather than treating adoption as the expected outcome. Previous studies have examined how children and their families use generative AI for storytelling, creative activities, and AI literacy \cite{Fan2024StoryPromptExploringThe,Druga2022FamilyAsA,Quan2025ParentsChildrenAnd}. Together, these studies show that generative AI can be used as a creative collaborator, a resource for knowledge- and play-driven conversations, and a shared subject of inquiry for children and parents. They also show that caregivers can help structure, guide, and make sense of children’s AI use.

To date, less is known about how parents evaluate generative AI as a participant in children’s pretend play. It remains unclear whether AI could have a role, what roles it could play, what risks these roles might introduce, and what requirements would make child--AI interactions safe and appropriate during the play. This gap motivates our study of how parents relate AI’s possible roles to children’s imagination, social relationships, and the conditions they see as necessary for safe and appropriate play in family environments.
\section{Methods}

\paragraph{Ethical Considerations, Data Collection and Participant Recruitment}

Approval from the Institutional Review Board of Drexel University was obtained for this study. Participants completed a short online screening survey to determine whether they were parents or legal guardians of children aged 4 to 15, fluent in English, residing in the USA, and able to complete an online survey and a Zoom interview. This age range was selected because pretend play is common in early childhood to middle childhood and may persist in different forms with considerable individual variations \cite{Smith2012PlayOnRetrospective}. Participants were recruited through convenience and snowball sampling between September 2025 and January 2026, and eligible participants were invited to a one-hour Zoom interview. Each participant received a \$20 online gift card as compensation. Participants received information on the study and were instructed about their options to withdraw from the study. All interviews were recorded with the participants’ consent, transcribed verbatim, cleaned for accurate analysis, and stored in a secure database accessible only to the research team.

Ten participants were recruited in total, and all recruited participants completed the survey and the interview. Before the interview, participants completed a brief survey used to confirm eligibility and then collect their demographic and household information.
Additionally, participants were asked to report on how frequently they used generative AI and rated seven concerns on five-point scales ranging from 1 (strongly disagree) to 5 (strongly agree) about AI fairness and bias, inaccurate information, personal data collection, data security, lack of inclusiveness, transparency, and oversight (Appendix \ref{app:survey-items}). We used the survey responses to verify eligibility and descriptively contextualize participants’ interview accounts; for a small purposive sample, participants represented a wide variance in the frequency of AI use.

\paragraph{Participant Characteristics}

The characteristics of the parents who completed the survey and interview (n=10, 100\%), including children's age ranges, parents' self-reported generative-AI use, and their concerns about AI are summarized in Table ~\ref{tab:participant_summary}. The participants consisted of 6 mothers and 4 fathers. Eight participants were aged 36 to 45 years old, while the other two were 26 to 35 years old. Educational attainment was relatively high, with 9 having completed graduate school. All participants reported having two children. Participants' children were aged 4 to 7 (youngest) and aged 8 to 12 (oldest) most commonly, covering early to middle childhood. Familiarity with generative AI varied across the participants: 2 participants used it daily, 2 weekly, 5 rarely, and 1 had never used it.  The complete participant-level responses to the survey are in Appendix Table~\ref{tab:participant-demographics}.

\begin{table}[t]
    \centering
    \caption{Summary of participant and child characteristics (\(N=10\)).}
    \label{tab:participant_summary}
    \small
    \renewcommand{\arraystretch}{1.15}
    \begin{tabularx}{\columnwidth}{
        @{}
        >{\raggedright\arraybackslash}p{0.27\columnwidth}
        >{\raggedright\arraybackslash}X
        @{}
    }
        \toprule
        \textbf{Characteristic} & \textbf{Sample summary} \\
        \midrule

        Parent age
        & 26--35 years: 2; 36--45 years: 8 \\

        Gender
        & Female: 6; Male: 4 \\

        Race/ethnicity
        & White/non-Hispanic: 5; Asian/non-Hispanic: 3; White, Asian, and Pacific Islander/non-Hispanic: 1; White/Hispanic: 1 \\

        Marital status
        & Married: 9; Divorced: 1 \\

        Educational attainment
        & Graduate or professional degree: 9; Associate or technical degree: 1 \\

        Annual household income
        & \$75,000--\$99,999: 2; \$100,000--\$149,999: 1; \$150,000 or more: 7 \\

        Profession
        & University faculty: 1; Attorney: 1; Product manager: 1; Clinical research manager: 1; Education consultant: 1; AI ethics researcher: 1; Primary school educator: 1; AI researcher: 1; Clinician: 1; Speech-language pathologist: 1 \\

        Number of children
        & Two children: 10 \\

        Youngest child's age
        & Under 4 years: 1; 4--7 years: 7; 8--12 years: 2 \\

        Oldest child's age
        & 4--7 years: 3; 8--12 years: 6; 12--15 years: 1 \\

        Familiarity with generative AI
        & Daily use: 2; Weekly use: 2; Rare use: 5; Never used: 1 \\

        \bottomrule
    \end{tabularx}

    \vspace{2pt}
    \begin{minipage}{\columnwidth}
        \footnotesize
        \textit{Note.} Demographic information and familiarity with generative AI were collected through the participant survey. Information about participants' professions was collected during the interviews. Participant-level characteristics are reported in Appendix Table~\ref{tab:participant-demographics}.
    \end{minipage}
\end{table}

\begin{figure}
    \centering
    \includegraphics[width=0.85\linewidth]{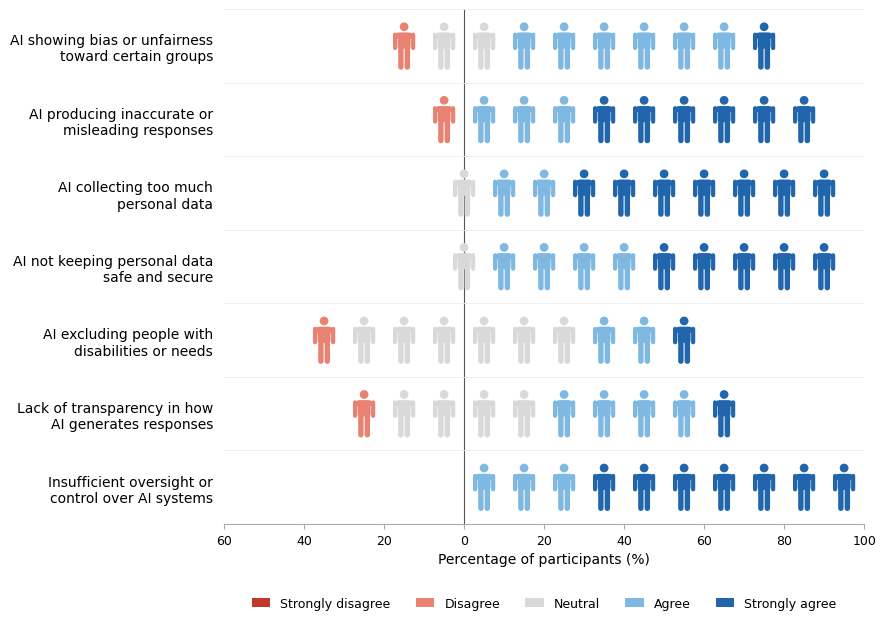}
    \caption{Participants’ Ratings of Concern About Potential AI-System Risks}
    \label{fig:Concern_Barchart}
    \Description{A diverging Likert-scale chart showing ten participants'
    levels of agreement with seven statements about potential AI system risks.
    Most participants agreed or strongly agreed that insufficient oversight,
    excessive personal-data collection, inaccurate responses, and inadequate
    data security were concerns.}
\end{figure}

As shown in Figure~\ref{fig:Concern_Barchart}, participants generally expressed concern about potential AI-system risks, particularly insufficient oversight or control and excessive personal-data collection. Participants’ ratings for all seven concerns are reported in Appendix Table~\ref{tab:ai_concerns}.

\paragraph{Interview Procedure}

We designed an interview guide to facilitate semi-structured interviews (Appendix \ref{app:interview-guide}). Each interview session was scheduled for 60 minutes and consisted of four different phases. Questions were informed partly by the Technology Acceptance Model~\cite{davis1989PerceivedUsefulnessPerceived}, the PEACE model of socio-emotional qualities in conversational agents~\cite{svikhnushina2022PEACEModelKey}, and Media Equation Theory~\cite{lee2008MediaEquationTheory}, which helped us examine parents' perceptions of the utilities, assumptions, constraints, the context of use, and relational potentials of technologies in children’s lives.

The initial phase of the interview focused on learning about the participants as parents and how they describe their children’s pretend play and their experiences, followed by how they evaluate their children’s current technology use, if any. We introduced a broad description of pretend play, acknowledging that imagination-based pretend play can also be present in other types of play, and it can be hard for parents to distinguish them based on their everyday observation. We did not ask the participants to focus on a specific child. So, when parents had multiple children within the eligible range, they were free to draw on their experiences and observations from any of their children.

In the second phase of the interview, we introduced a short narrative describing the AI in children’s pretend play concept to the participants. We shared the screen view of the description via Zoom so the participants could read the description as one researcher read it aloud. We introduced the AI in children’s pretend play concept by referring to it as an AI chatbot without prescribing a detailed form factor, functionality, use cases, fidelity, implementation, or other technical features. This open framing allowed participants to create their own mental model of the system and children's and parents' experiences based on their different parental characteristics, preferences in technology, as well as their current views of AI at the time of the interviews. Thus, an AI chatbot in this context could have ranged from conventional chatbots, conversational agents, natural-language interaction, machine-learning and/or natural language processing (NLP) algorithms, generative AI, and multi-modal AI systems. This way, we also invited participants' views on AI and AI for children in a broader context. Parents then shared their thoughts on AI in children’s pretend play by answering our questions, which revealed their evaluation of our concept as well as their broader orientations toward AI and AI for children. 

In the third phase of the interview, we showed storyboards illustrating two boundary cases of child-AI interaction (Appendix \ref{app:scenarios}). These had been prepared in advance by two researchers through affinity diagramming and categorization of potential risks identified in previous studies~\cite{jiao2025LLMsChildhoodSafetya, khoo2025MinorBenchHandbuiltBenchmark, rath2025LLMSafetyChildren}. Follow-up questions were asked to probe participants’ interpretations, concerns, and design expectations in response to these scenarios. 

\paragraph{Data Analysis and Synthesis}

We used a combined deductive and inductive approach. First, we conducted a codebook thematic analysis~\cite{braun2006UsingThematicAnalysis, Braun2023TowardGoodPractice, Roberts2019AttemptingRigourAnd}. Two researchers first conducted open coding to develop a codebook using an inductive approach, labeling and defining recurrent patterns in all transcripts as codes during weekly meetings. Discrepancies were resolved through discussion, and if necessary, a third researcher facilitated consensus. The codebook has been iteratively refined with the input of the last author. This process was iterative until we finalized the codebook and fully coded all transcripts. Coding was managed collaboratively in a shared Excel spreadsheet; the finalized codebook is presented in Appendix \ref{app:codebook}.

\begin{figure*}[t]
    \centering

    \begin{subfigure}[t]{0.48\linewidth}
        \centering
        \includegraphics[width=\linewidth]{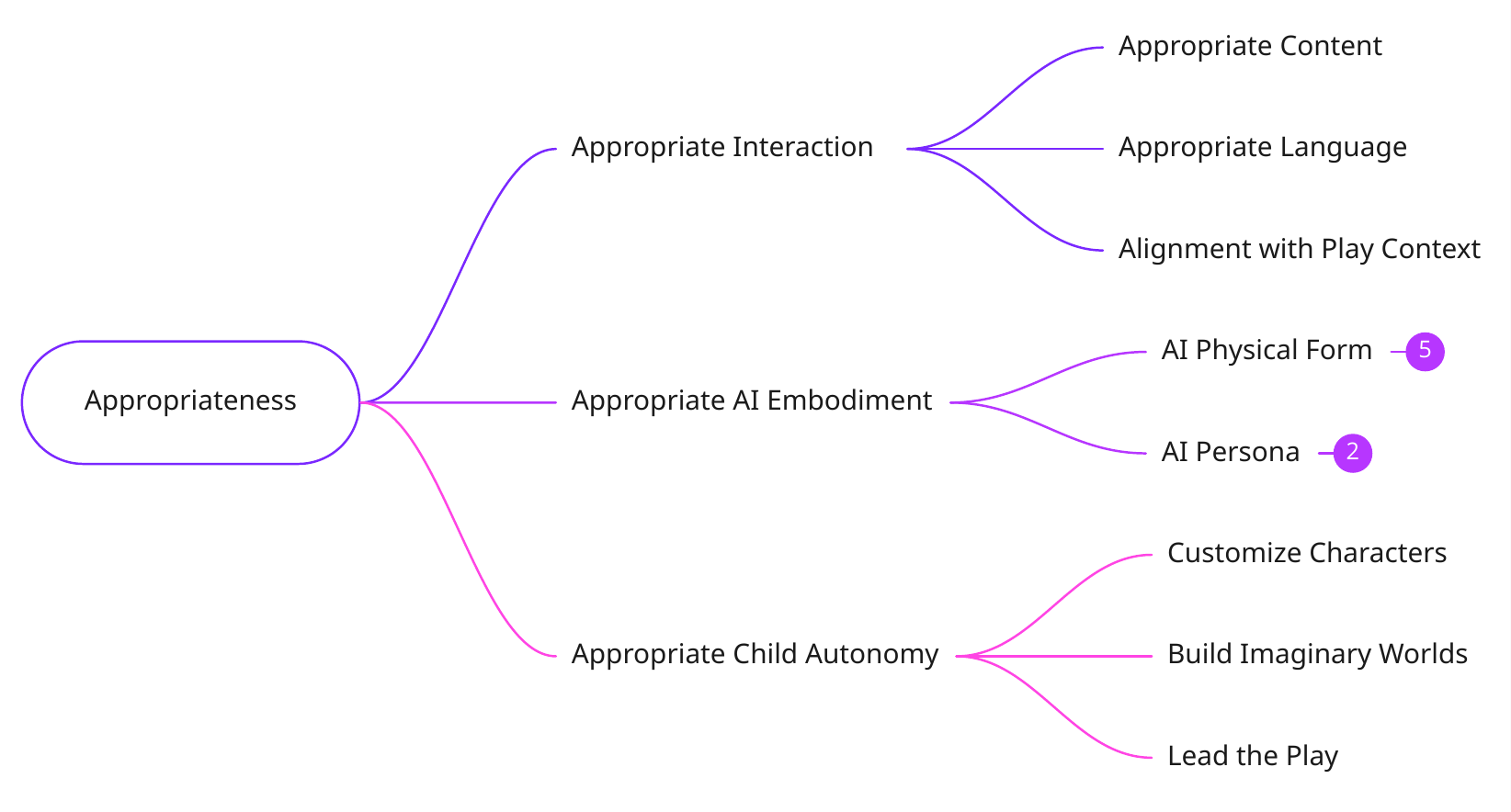}
        \caption{Parents' views about appropriate features and interactions in AI-supported pretend play: Concept Map.}
        \label{fig:appropriate-a}
    \end{subfigure}
    \hfill
    \begin{subfigure}[t]{0.48\linewidth}
        \centering
        \includegraphics[width=\linewidth]{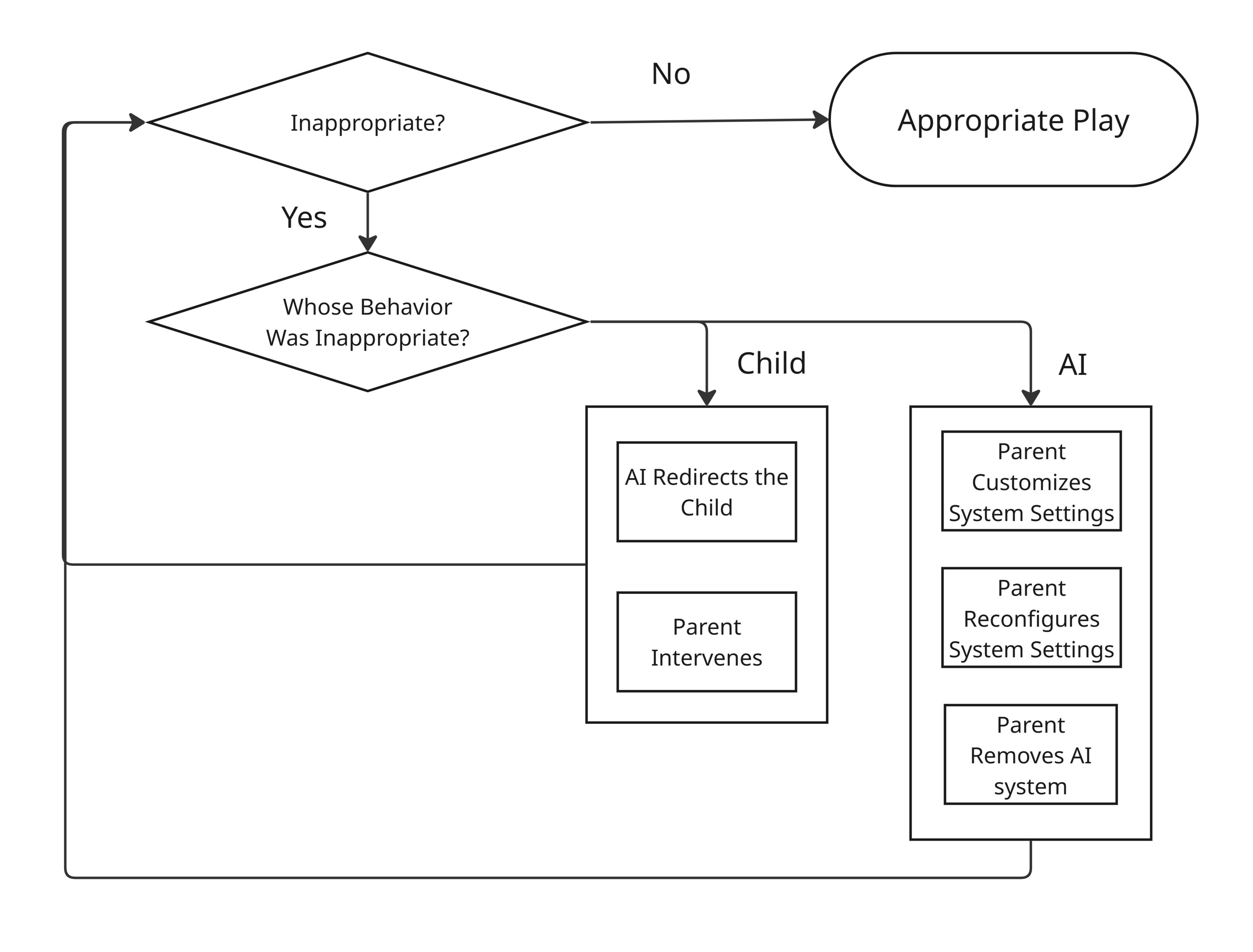}
        \caption{Parents' conception of how inappropriate interaction could be handled: Feedback-loop diagram.}
        \label{fig:appropriate-b}
    \end{subfigure}

    \caption{Examples of Systems Diagrams Used to Understand the Complexity of Parents' Perceptions.}
    \Description{Two horizontally arranged systems diagrams. The left diagram maps parents' views of appropriate features and interactions in AI-supported pretend play. The right diagram shows a feedback loop for responding to inappropriate interactions and maintaining appropriate play.}
    \label{fig:systems_diagrams}
\end{figure*}

After developing the codes from participants' responses, the two researchers conducted affinity diagramming on a Miro board: the coding was organized deductively around the research questions and related codes were clustered into initial themes. To develop a more holistic understanding of the data, we created qualitative systems maps \cite{Hanger-Kopp2024WhatQualitativeSystems}. These maps helped us explore how concepts, conditions, and possible mechanisms articulated across participants' accounts related to one another. Rather than serving as direct maps of the codebook or final themes, they were internal analytic representations that supported theme refinement and cross-thematic synthesis (Figure \ref{fig:9CrossThematic}). Figures~\ref{fig:appropriate-a} and \ref{fig:appropriate-b} illustrate examples of two types of diagrams used during this process: a hybrid cognitive--concept map and a feedback-loop diagram. These figures are illustrative analytic representations, not stand-alone empirical models or additional findings.
\section{Results}

\subsection{RQ1: What Are Parents' Perceptions about the Potential Roles of AI in Children's Pretend Play?}

\subsubsection{AI as an Accessible, Child-Directed Play Partner}

The first theme captures how parents imagined AI as an accessible, child-directed partner in role-based social pretend play. Parents imagined AI as an accessible role-play partner that could assume characters, respond to imaginative directions of children, and maintain social pretend play.

When describing their involvement in their children’s pretend play, parents discussed supporting play activities by providing play objects, presenting ideas for more creative play, co-playing, and facilitating peer play. In alignment with these existing practices, parents evaluated the AI in children's pretend play concept for its potential to scaffold pretend play and extend its developmental benefits---AI was not just another source of child entertainment to the parents. Thinking of its functional capacity, parents imagined AI could be added as a play object or a play partner to enrich children's play in the future (n=6, 60\%). P5 anticipated that AI should create highly engaging experiences for children by fostering imagination and creativity:

\begin{quote} ~\textit{"I could see benefits. [...] I think it could be very engaging. [...] Having a device that [...] allows my kids to practice being imaginative and creative, that's wonderful."} \end{quote} 

Here, engagement was not a sufficient indicator of quality pretend play acceptable to the parents, as parents’ accounts frequently deviated from using AI as a parental aid to occupy children's time and attention (n=7, 70\%). P8 favored children's agency in using their imagination during play time:

\begin{quote} ~\textit{"I think creativity is so important, and imagination also. Kids being bored is really important for them to find time to come out and create new worlds or create problems to solve. [...] I like watching movies with them, but I think the imagination part is really so important."} \end{quote} 

This presents a requirement for AI systems to scaffold children's creative play without taking their imagination and agency away so they can continue to lead and sustain pretend play. In summary, if AI is to be used to enhance children’s pretend play, parents' focus is not on entertainment alone. They focus on its functionality to extend developmental benefits of pretend play by improving the quality of play that does not take away children’s agency over their imagination. 

In support of children's social development, some parents suggested providing social training for their children in a role-play format or modulating children's behaviors through dialog-based child-AI interactions (n=4, 40\%). One example was a scenario in which the child would learn to navigate interpersonal conflicts. 

\begin{quote} ~\textit{"It would be a nice option for children who have a difficult time with socialization [...]. Maybe [...] have the AI chatbot have healthy disagreements with your child to learn how to deal with conflict."}~(P7) \end{quote}  

In this case, P7 viewed AI having disagreements with the child as opportunities for children to rehearse emotional regulation, perspective-taking, and conflict resolution. Remarkably, engagement in pretend play may differ across children as well as the opportunities for social growth it affords. Remarkably, the expression of pretend play may vary across children. P7 suggested that AI could help by ~\textit{"kind of supporting my child who doesn't pretend play as naturally."} 

According to P1, AI could provide neurodivergent children with play experiences individualized and aligned with their different communication styles in meeting their developmental needs: 

\begin{quote} ~\textit{"It would be interesting to see him use it as a social situational primer before he has to go out and meet other kids. He can get practice with this chatbot, learning how to converse [...]. So, if a child has some kind of social communication problem, then that could be a good tool."} \end{quote}

In summary, parents envisioned the social components (e.g., role-based interaction) of pretend play to be carefully designed to support children who may experience social anxiety or impairments. 

\subsubsection{AI as Assistants for Parents}   

Parents hoped that AI could help them co-play with their children or guide their children’s pretend play. With caregiving fatigue as a reality for some parents, playing with small children could be an additional burden. Parents imagined AI as a parental aid that could alleviate this burden (n=7, 70\%); for example, by occupying their children when they needed to mind other responsibilities (n=4, 40\%). One of the parents put this idea in the following: 

\begin{quote} ~\textit{"[...] a parent aid where we can actually try to get something done, because kids are always wanting your attention’’} (P4). \end{quote}

Most of these parents thought that it would be better than administering screen time to their children, describing it as \textit{``a better alternative than turning on YouTube and having them watch whatever’’}~(P1). 

Ultimately, parents saw AI assisting with children's pretend play as a way to aid parental responsibilities and as a better alternative to traditional screen time. However, this thought subsequently led them to worry about the potential negative effects of overreliance on AI for convenience. 

Some parents reported that playing with small children could sometimes feel mentally exhausting (n=3, 30\%). P6 recollected her previous experience doing pretend play with her children: 

\begin{quote} ~\textit{"It was just a lot of mental energy to put myself in the mind of a child, playing pretend with my older daughter"}. \end{quote}

The reason is simple; it is difficult for adults to match children's level in play. P5 described this as:

\begin{quote} ~\textit{"It's not as natural to me, because it required so much creative imagination and silliness. [...] They would want me to---and I would do it sometimes---but I tried to get out of it. [...] I don't know if everybody finds it as hard as I do."}. \end{quote}

Thus, continuing to fill the play with imagination and creativity to meet the expectations of small children was found to be challenging for adults.  

In this matter, parents pictured AI making co-play easier, which could increase parent-child playtime (n=5, 50\%). Based on the parents' accounts, there were three ways for AI to do this. First, AI could help parents who are observers of children's pretend play. P6 imagined generative AI as a source of creative ideas for parents to use to scaffold children’s imagination during play:

\begin{quote} ~\textit{"There’re a lot of instances where we actually want to do it ourselves [...], but when you do pretend play, you need generative machinery"}. \end{quote}

Thus, AI could be used to support the creation of new characters, imagined environments, or ways to interact with the characters. In this case, the parent remains physically present for the child without direct participation in the play, overseeing and mediating child-AI interaction to safely scaffold children's autonomous exploration (i.e., initiating the play and continuing to lead it). There were some other parents who wanted to be more involved in co-playing with their children, but wanted AI to make participation easier. P4 expressed:

\begin{quote} \textit{"Maybe the parent’s like, “I'm not creative or something, and I need help telling a story.”"} \end{quote}

The same parent requested AI to help them come up with a story so that both the parent and the child could fill in the details together. Here, AI could help them continue narrating their ideas by generating artifacts (e.g., illustrations) if they get stuck and are not able to move forward. In this case, parents would interact directly with their children, and AI would be mediating their interaction whenever needed. 
The other parents imagined themselves playing a pretend-play family game that consisted of imaginative exploration. P8 described that more human players, i.e., the child, the parent, and AI-generated characters, would interact with one another in the imagined world created by the child: 

\begin{quote} \textit{"[...] if you have two kids [...] they're both playing with it, or if them and my wife are playing with it."} \end{quote} 

In such a pretend-play family game, children's interactions with other human and AI players would be guided by the parent. In summary, parents desired that AI should make pretend play with their children easier and more enjoyable. Additionally, parents imagined AI in children's pretend play to assume the roles of a play partner and/or characters in the play, further enhancing the quality of play experiences for their children. 

\subsection{RQ2: What Types of Risks Did Parents Associate with Children Engaging in Pretend Play with AI?} 

\subsubsection{AI as a Threat to Human Relationships}

Parents feared that children’s engagement with AI could weaken parent--child and other human relationships, with these concerns intensifying when they imagined children forming emotional attachments to AI (n=9, 90\%). They stated that children bonding with technology emotionally should not be the norm. P5 described feelings against this thought:

\begin{quote} ~\textit{"To develop an emotional attachment to a device, an algorithm [...] is very strange and new and different, so there's a lot of unknown dangers in that, when it's just a bunch of ones and zeros."} \end{quote}

Another risk factor that concerned the parents substantially was reduced human connection, identified as reduced human-to-human interaction and the possibility of AI replacing the roles of important people in children’s lives. They worried that child–AI interaction during pretend play would compete against human interaction for children (n=8, 80\%). This was particularly concerning because human interaction was regarded as indispensable for children: 

\begin{quote} ~\textit{"In the long run, they're going to have to interact with other people."}~(P2). \end{quote}

Because child–AI interaction differs from natural human-to-human interaction, AI may not support children's social development in the same way:

\begin{quote} ~\textit{"You're never gonna have a conflict with your AI robot. They'll just do anything you say."}~(P2). \end{quote}

Thus, parents felt that children need experiential learning opportunities, e.g., naturally occurring social conflicts. Parents particularly emphasized learning empathy, which they thought AI would not be able to support in the same way:

\begin{quote} ~\textit{"They [iPad kids] just have a hard time sympathizing, empathizing with other people [...] Obviously, this is not all from AI [...], but I could just see where it could exacerbate the issues."}~(P10). \end{quote}

Thus, parents perceived that child–AI interaction could negatively affect children's communication, social, and emotional development by altering typical developmental patterns. In effect, the benefits that role-based social pretend play is believed to provide may be compromised by these limitations of AI. 

Moreover, the thought of their children interacting with AI increased their concerns over AI potentially taking the roles of important humans in children's lives (n=8, 80\%), leading to compromised relationships with other people:

\begin{quote} ~\textit{"The main thing that would make me pause is if my child was forming a stronger relationship with the technology than with other people."}~(P6) \end{quote}

This worry extended to the weakening of parent-child bonding due to the uncertainties that could be introduced into the family dynamics:

\begin{quote} ~\textit{"We want our children to share their thoughts with us as well. [...] Having the robot be the best friend is what we would want to avoid."}~(P4). \end{quote}

Such a relational threat could become more severe for companion-like AI compared to a tool-based AI. To mitigate these concerns, parents wanted AI in pretend play to remain within clearly defined functional boundaries; supplement human play partners rather than replace them. Another reason why AI should not replace human roles is its technical and functional limitations. P10 described an important limitation of AI\hypertarget{p10_nonverbal}:

\begin{quote} ~\textit{"It's not gonna express some of those subtle nonverbal communications that we would exhibit. [...] As someone that's working with language-impaired children, I feel like I've been seeing more and more of kids where they just don't even know how to use context clues, even with reading."}. \end{quote}

Based on her experience, P10 speculated that children's AI use would impair children's development of language and communication skills, because AI could not reproduce unique qualities of human communication underpinning children's social and language development, such as nonverbal, subtle, and contextual cues. 

Overall, all parents were aware of potential risks of AI for children. They viewed emotional attachment to AI as a plausible and prominent compounding risk factor relating to children's pretend play; this could aggravate other risk factors such as reduced human-to-human interaction and replacing the roles of important humans in children's lives, both weakening parent-child relationships. These risks would not only impact user experience negatively, but also relate to long-term negative consequences in children's communication, social, and emotional development. 

\subsubsection{Parental Responsibilities as a Boundary Setter}

For many parents, parenting is already a complex task because growing children change and present new challenges every day. They described parenting as a continuous act of balancing children's need for protection and autonomy:

\begin{quote} ~\textit{"As a parent, I think it's my job to protect them [...] from things that are harmful."}~(P5) \end{quote}

At the same time, parents recognized that children needed increasing autonomy to develop their own problem-solving abilities:

\begin{quote} ~\textit{"It's important for the kids to grow their own muscles [...] so that they could handle situations themselves."}~(P7) \end{quote} 

Parents stated that children's technology use would create new roles for them (n=8, 80\%) as a "supervisor"~(P6) who guides children's technology use as well as a gatekeeper and a monitor~(P5), which was reflected in the following accounts: ~\textit{"I think parents should be 100\% supervising, [...] even just internet use."}~(P6) to ensure healthy technology relationships. Thus, AI introduced another dimension to parents’ ongoing negotiation between protecting children and supporting their autonomy, while creating new responsibilities to understand, evaluate, and potentially participate in their children’s AI use.

Parents’ sense-making of their role in their children’s potential use of pretend-play AI reflected three main stances: an investigative stance (n=2, 20\%), a participatory stance (n=3, 30\%), and a generally negative stance (n=5, 50\%). An overall negative stance means that parents were against their children using AI; an investigative stance means that parents wanted to learn about how the AI works to ensure safety; and a participatory stance means that parents expressed the desire to participate in pretend play with their child and the AI.  
 
First, imagining AI introduced into children’s pretend play, some parents thought of new parental duties related to children’s technology use that go way beyond current practice. These included teaching their children about AI, guiding their AI use, stepping in as a parent if needed during or after play, and tailoring/testing the AI system themselves. To carry out these new tasks, some parents took an investigative stance toward AI, requesting information about security features, the goals of AI, and how the system operates:

\begin{quote} ~\textit{"I would really deeply want to understand the security around this to make sure it's just a dragon leading through a fantasy land and can't turn into anything worse than that."}~(P8). \end{quote}

Some of the investigative parents also stated that they would teach their children about AI themselves:

\begin{quote} ~\textit{"I'll very much tell them how it works and that it can't be trusted, not in a way to scare them, but just like, ‘This is not a person. This is not your teacher,’ or ‘This is not a human being that's thinking about reflecting on what it's saying.’"}. \end{quote}

These were examples of additional parental responsibilities that they anticipated. P6 described this new parenting role clearly in the context of guiding appropriate technology use: 

\begin{quote} ~\textit{"Those Leapstart books, where you press and then it tells you the word. It's supposed to teach you reading. My kids never use those things the way they're meant to be used. They always turn it into---they're just getting some dopamine from hitting the buttons. But, if I go sit with them and then we talk about it, [...] So, I would probably do a similar thing with this [AI for pretend play]."}. \end{quote}

Thus, parents play an important role in guiding children’s AI use to support their well-being at home. However, because not all parents feel prepared or willing to explore AI tools independently, AI designers should provide parents with accessible information and tools for understanding and mediating their children’s AI use.

Second, parents felt responsible for increasing their participation in pretend play:

\begin{quote} ~\textit{"There's part of me that feels like a guilty parent. ‘Well, I should be doing that [pretend play with my child]. I'm not doing my job if there's an AI doing it instead?’"}~(P8). \end{quote}

Parents disallowed AI from completely taking over the role of playing with their children, and they (n=3, 30\%) conceived a pretend-play family game:

\begin{quote} ~\textit{"If the whole family was playing make-believe and we added a chatbot, that could be interesting and exciting."}~(P8) \end{quote}

This view may depend on whether they were already playing with their children:

\begin{quote} ~\textit{"I think I'd enjoy using it with them. [...] maybe it's just my style as a parent."}~(P6). \end{quote} 

The idea of a family pretend game might help increase parent-child playtime, which may create an opportunity to learn better about their children, to teach them appropriate ways to use AI, or to confirm AI behaving within the pre-configured boundaries. P8 described the importance of AI staying within an imaginary world:

\begin{quote} ~\textit{"In pretend play, [...] it's your imaginary friend. [...] It’s a partner that you build these worlds with. That partner is very important in the artificial world, but that partner is not helping you decide what to do for school."}~(P8). \end{quote}

These accounts from the parents with a participatory stance toward AI-supported pretend play suggested the idea of a pretend-play family game. Regardless of the viability of such an approach, the idea that it can provide engaging playtime for both parents and children, while minimizing AI risks, exhibits significant conceptual merit. 

Lastly, some parents held an overall negative stance toward AI because they feared potential negative outcomes. Due to the uncertainties about AI's influence on children, P2 feared any irreversible damage on his child:

\begin{quote} ~\textit{"You just don't know what the results are gonna be, and you can't take them back. [...] There's no real guardrails or regulations on this thing."} \end{quote}

When comparing AI with other technologies, parents framed the risks of AI as uncertain risks, because AI is new, constantly evolving, and not validated yet:

\begin{quote} ~\textit{"When I use the word, 'risk for AI,' I'm referring to a lot of unknown dangers, [...] because AI is so untested [...] and changing really fast."}~(P5). \end{quote}

For this, parents felt that they did not have the knowledge about AI required to keep their children safe:

\begin{quote} ~\textit{"Oh, shoot! I don't know what this thing [AI] is necessarily. [...] I'm not an engineer.”}~(P2) \end{quote}

These accounts illustrate the magnitude of parents' concerns about AI for children. Parents perceived AI as unpredictable and likely causing long-term negative influences on their children. They believed that safety measures such as regulations and safety guardrails were insufficient. In summary, parents anticipated that AI in children's pretend play would create new parenting tasks, where different parents would have different capacities and information needs.   

\subsection{RQ3: How do Parents Imagine a Safe and Appropriate AI Design for Pretend Play?}
    
\subsubsection{Parental Protection for Child Safety}

Assuming that safe AI-supported pretend play could be achieved, we explored what parents believed an AI system would need to prevent potential harms for children during pretend play with AI and how they would protect their children if allowing such child–AI interaction. First, parents shared their thoughts about how they would evaluate AI systems for safety with the awareness of potential dangers associated with AI for children. As a baseline requirement for safety, parents requested robust safety guardrails embedded in the system (n=7, 70\%):

\begin{quote} ~\textit{"With children, you gotta put those guardrails up really high and make sure that [...] it's as safe as it could possibly be."}~(P1) \end{quote}

Here, parents treated safety guardrails as a prerequisite for system use, suggesting that a system perceived as potentially unsafe would not be acceptable for their children. Next, parents needed clear information about AI's intention and its operation:

\begin{quote} ~\textit{"What are the goals of the AI [...] or the interaction for the child?"}~(P4). \end{quote}

In addition to the system-level guarantees, parents requested the validation of long-term effects on children as well as signs of public acceptance:

\begin{quote} ~\textit{"For technologists and child psychologists to understand the effects of AI on people and children. [...] If it's more like a social norm [...], because the kids love it and so many people are using it, and [...] they're not troubled nor having breakdowns."}~(P5) \end{quote}

Notably, safety assurance alone was insufficient for earning parents’ full trust. After considering ways to implement system guardrails, P10 explained how AI might never gain her full trust as a parent, reflecting her continued vigilance in protecting her child:

\begin{quote} ~\textit{"I guess like guardrails on topics, maybe? Trying to think. [...] I don't want to sound cynical. The problem is [...] it would probably never have my full trust, but I wouldn't put my full trust into pretty much almost anything"}. \end{quote}

Thus, parents' unwillingness to place trust in AI for children seems to extend beyond the distrust of the technology and reflect a broader parental vigilance toward anything that could influence their children.

Parents believed that AI is constantly evolving and adapting, posing uncertain dangers that complicate family dynamics and parenting. Consequently, they felt they had to actively ensure its safety for their children. (n=9, 90\%), and parental control was of the utmost importance to them:

\begin{quote} ~\textit{"I would never, absolutely never ever give them unrestricted access to converse with AI."}~(P6). \end{quote}

P3 suggested several areas of parental involvement for child safety:

\begin{quote} ~\textit{"Parent customization and parent control would make me feel pretty safe. [...] Being there for the initial interactions and then also monitoring play over time."} \end{quote}

Here, we learned that parental protection would be a multi-layered process, and parents envisioned themselves having to do the work to make it safe for their children by configuring the system, setting parental controls, and monitoring interactions. 

As for parents configuring the system, P7 listed possible parameters in the settings:

\begin{quote} ~\textit{“What kind of characters they can be, the situation they might be putting themselves in, [...] how long they're able to have their interactions for, making sure that the dialogue goes back and forth, […] not one versus the other is dominating the situation or the pretend play.”} \end{quote}

Considering each child’s unique needs, P1 recognized that safety configuration thresholds would be personalized and vary for every individual:

\begin{quote} ~\textit{"Kids are all different, [...], so it has to be able to be customized to work with different kids and what their interests are. [...] I would do it for my youngest. [...] I wouldn't want to create something that would be used by a bunch of people."} \end{quote}

This raises a question about whether the system could accommodate individualized safety guardrails and interaction styles. 

Ideally, parents should be present to be able to handle unpredictable situations that may occur during child-AI interactions. Otherwise, the need for robust parental controls would become paramount:

\begin{quote} ~\textit{"You gotta be there as a parent. If you can't be there as a parent, you gotta have strong controls over what they can and cannot access."}~(P2). \end{quote}

While a time limit and content filters were two main areas of parental control, additional needs for interaction control were also brought up. Regarding dialogue-based AI interactions, parents preferred that AI-child dialogue remain within the chosen storyline. However, it remained unclear whether children should be allowed to have open-ended, natural conversations. For P1, this decision depended on several factors, including the purpose of AI interaction:

\begin{quote} ~\textit{"If you want a more natural conversation, that would be the way to go, but [...] if I'm trying to teach my kid to be social and practice conversing with others, maybe bringing it back. [...] It depends on the kid. Depends on what you want."} \end{quote}

Thus, parents should be able to adjust the degree of freedom of child-AI interaction, and the system should redirect the child user back to pre-selected dialogue as needed to remain within safe boundaries. 

To help manage safe child-AI interactions, parents proposed a notification feature. P3 wanted to be notified if their child moved beyond the configured pretend-play interaction and initiated unconstrained dialogue with the AI:

\begin{quote} ~\textit{"If they want to venture up, then maybe that will be like an alert to the parents."} \end{quote} 

Other types of information that parents desired to be notified of were what storyline and play characters the child created as part of the AI-supported play. Additionally, P3 wanted to be notified of any atypical emotional responses from the child expressed during the play:

\begin{quote} ~\textit{"If they had any negative sentiments or overwhelmingly positive sentiments, I want to know […], I wanna know why."} \end{quote}

These requirements would necessitate a monitoring capability for timely detection and notification:

\begin{quote} ~\textit{“I probably also want some sort of monitoring function […], so I can be assured that the conversations are appropriate.”}~(P5). \end{quote}

To ensure long-term safety, parents expected to receive regular interaction summaries in a newsletter. For parents, continuous monitoring also meant an opportunity to calibrate the system if necessary. P6 conceived a way that parents could modulate AI responses to be aligned with the child's age:

\begin{quote} ~\textit{"[...] sliders so that you can understand what the model thinks about you, and you can also calibrate it [...]. It can’t suddenly think that they’re an adult."} \end{quote}

In summary, parental perceptions revealed that safe child-AI interaction in pretend play was multi-layered. Beyond a baseline of robust system-level safeguards and external validation, parents want the technology to facilitate parental monitoring to ensure safety and allow calibration to accommodate the changing needs of maturing children.

\subsubsection{Appropriate AI for Children’s Play}

In parents' perceptions, safety was a necessary but not a sufficient condition. Parents framed additional requirements against the notion of 'appropriateness' (n=10, 100\%). These dimensions of appropriateness were discussed: linguistic/narrative appropriateness, situational appropriateness, and embodiment appropriateness.

First, the linguistic/narrative appropriateness of AI responses concerned what AI would say and how during pretend play, such as acceptable language, age-appropriate topics and content, words that children could understand, and coherence with the unfolding imaginary narrative. Parents mentioned age-appropriate play topics and content:

\begin{quote} ~\textit{"If you thought it [AI] was not supposed to discuss xyz or things that aren't age-appropriate."}~(P4) \end{quote}

P9 characterized appropriate interaction in terms of boundaries around acceptable language use:~\textit{"Age-appropriateness, cause I don't even let her say the fake curse words."} Additionally, AI responses should match the child's comprehension level:~\textit{"I guess you just gotta find a kid's way to say it."} Beyond discretionary choices of individual words and topics, P8 expected the AI’s responses and behaviors to remain coherent with the unfolding context of the pretend play storyline:

\begin{quote} ~\textit{"If this was more a companion chatbot, I would expect it to be happy, sad, or excited [...]. If it's strictly a role-playing chatbot, then I wouldn't expect that unless, in the adventure they're in, the chatbot got hurt, and then [...] they're sad. If it was context-appropriate, I think it would be okay. If they were trying to create a world and the chatbot just came across as depressed and then took them out of the game, then I think that would be different.”} \end{quote}

In short, narrative and linguistic appropriateness reflected parents’ expectation that the AI should communicate in ways appropriate for children while remaining coherent with the evolving context of the imagined storyline.

Appropriateness also referred to social situations during role-based pretend play. P7 described an imaginary social conflict that could become inappropriate:

\begin{quote} ~\textit{"It's okay for her to work out the fight between the Barbies. I guess it's nice when I am nearby, and I hear if the fight's getting into inappropriate territory."} \end{quote}

Furthermore, parents warned that the AI must never misbehave toward the child:

\begin{quote} ~\textit{"If they were trying to create a world and the chatbot just came across as depressed and then took them [children] out of the game, then I think that would be different."}~(P8) \end{quote}

Instead, they expected AI to maintain a positive tone:

\begin{quote} ~\textit{"If they were using naughty language or something [...], I wouldn't mind it redirecting its behavior a little bit in a positive way."}~(P7) \end{quote}

Thus, parents wanted the AI play partner to help children handle social conflicts constructively, while avoiding negative behaviors and promoting a positive, supportive tone.

Finally, a high-fidelity human-like embodiment was thought of as inappropriate, as it evoked unsettling emotions:

\begin{quote} ~\textit{"Trying too hard to look real, it's creepy. I imagine for the kids it's the same thing."}~(P6) \end{quote}

Children developing strong emotional attachment to the human-like AI character was another concern:

\begin{quote} ~\textit{"With stuffies and dolls, you do naturally grow out of it. This is so human-like. It makes me worried that they would not grow out of this."}~(P7) \end{quote}

Parents also shared their views of appropriate personas for AI characters; there were slight differences in their preferences. For example, P9 preferred a peer persona:

\begin{quote} ~\textit{"There's a difference between Bratz and Barbies. You know what I mean? Barbie is a very, very grown-up woman. So, I don't want my kid playing with a woman, I guess."} \end{quote}

In contrast, P6 leaned toward a responsible adult figure who is favored by children:

\begin{quote} ~\textit{"There are a lot of characters on media that have played this role of doing pretend play with kids really, really well. [...] people that have a talent for knowing how to play pretend with kids in an engaging and non-creepy way."} \end{quote}

P8 suggested that children should be able to customize AI characters for their play:

\begin{quote} ~\textit{"I think it would be fun for the kids to be able to customize […] to get the different accents and characters."} \end{quote}

Thus, to avoid uncanny realism and satisfy diverse persona preferences, parents suggested a collaborative design process where they can review and approve their children's customized AI characters. In summary, parents evaluated multiple components of AI pretend play for appropriateness, focusing specifically on narrative storylines, social interaction styles, and character personas. These environmental and behavioral requirements were viewed as imperative, although specific thresholds and preferences varied among parents.

\subsection{Cross-Thematic Synthesis}

\begin{figure}
    \centering
    \includegraphics[width=0.90\linewidth]{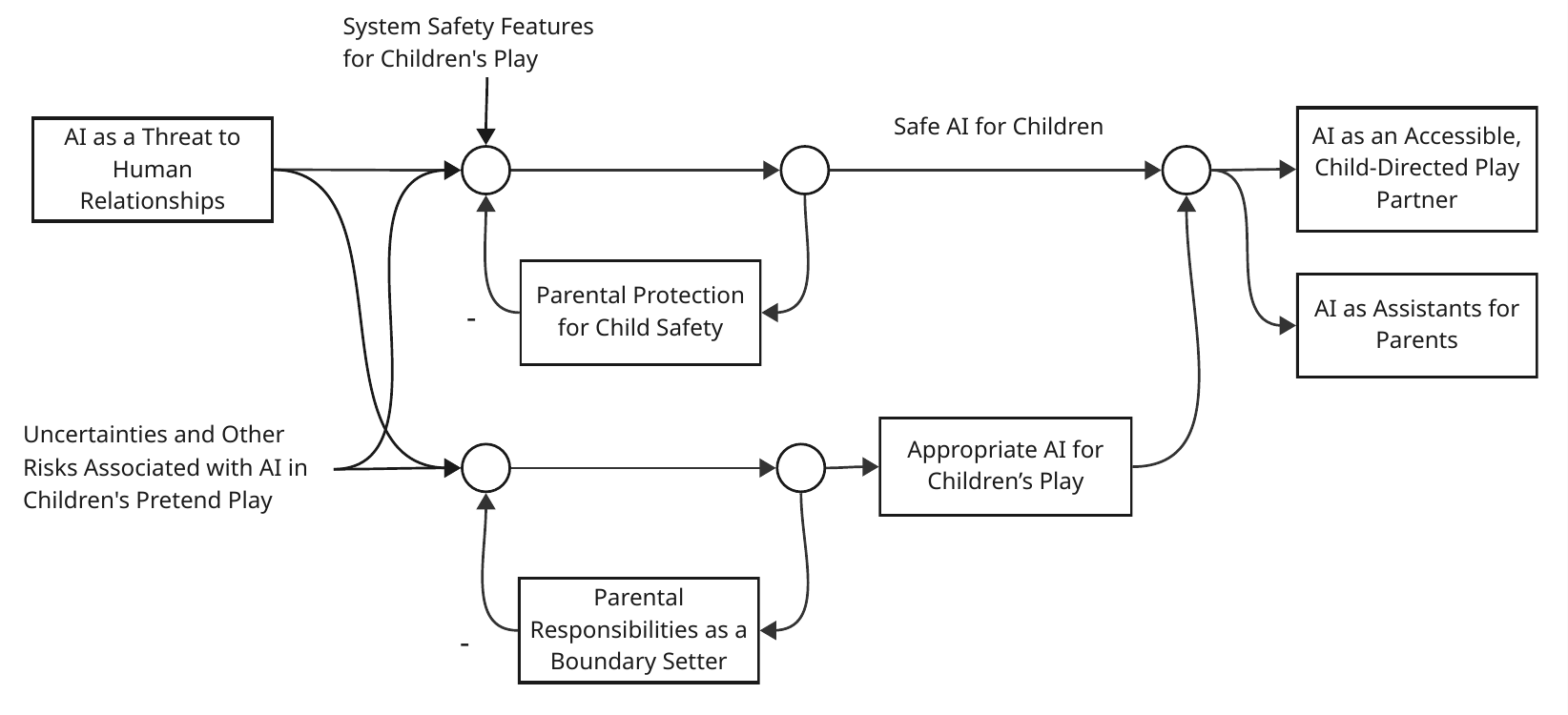}
    \caption{Cross-thematic Relationships among Perceived Risks, Safeguards, Appropriateness, and Potential Roles of AI.}
    \Description{A cross-thematic systems diagram showing how parents connected perceived AI risks, parental safeguards, design appropriateness, and AI's potential roles in children's pretend play. Perceived risks, including threats to human relationships, inform requirements for appropriate AI. Parental boundary-setting and protection are shown as ongoing safeguards rather than as guarantees of appropriateness. Appropriate, accessible, and child-directed AI play may enable AI to serve as a play partner for children or as an assistant to parents.}
    \label{fig:9CrossThematic}
\end{figure}

Figure~\ref{fig:9CrossThematic} brings the themes together by showing safety and appropriateness as related but distinct conditions for AI-supported pretend play. Parents' accounts suggest that system safety features and parental protection would both be needed to manage risks to children, while determining appropriateness would require continuing attention to relational concerns, other uncertainties, and boundaries specific to the child and family. Neither system safeguards nor parental involvement alone would guarantee acceptable child--AI interaction; parents anticipated an ongoing process of checking, limiting, and adjusting how AI participated in play. Only when AI could be considered both safe and appropriate did parents envision its possible benefits as an accessible, child-directed play partner or an assistant for parent--child play. Thus, the potential value of AI in pretend play remained conditional on shared responsibilities among system designers, providers, and families. The diagram represents an interpretive synthesis of relationships across the themes rather than a tested causal model.

\section{Discussion}

This study examined how parents imagined the possible roles, risks, and design requirements of AI in children’s pretend play. Parents saw value in AI-supported play, but they expressed concerns that children could become emotionally attached to AI, potentially affecting parent--child relationships and reducing human interactions necessary for children’s development. Additionally, their accounts indicated that making AI safe for children involves more than filtering harmful content. Parents shared their thoughts on how AI should behave in a pretend-play storyline, what social role it could take in the family, and what work parents would need to do to keep children's interactions with it safe and appropriate. 

This study contributes to understanding parents' sense-making of a speculative design for AI-supported pretend play and how they articulated their concrete expectations, concerns, and conditions for its use. In this study, we found not only their requirements for safe and appropriate AI, but also an existing frustration that pretend play can be difficult for some children and a desire for a pretend family game that AI may support to strengthen parent--child interaction. 

We discuss six implications based on our findings. The first two concern children’s agency in imaginative play and accessibility in social role play, and the next two focus on mediated AI interactions and AI’s place in the family ecosystem. The last two address parental protection for AI safety and the conditions under which AI could be considered appropriate for children’s pretend play. Table~\ref{tab:design-implications} lists the six implications at the discussion level for the corresponding design directions and priorities for future research. These were not part of our analytical findings, but reasoned inferences based on the findings. 

\subsection{Preserving Children’s Agency in Imagination and Creativity} 

Parents’ accounts point to one basic requirement: the child should have autonomy in leading the play and not outsource the imaginative cognitive work to the AI. For parents, children’s engagement should not be the only definitive characteristic for the quality of play. They saw children creating characters, deciding what happens, solving fictitious problems in the imagined world, and switching to a different storyline when they wished. AI could support these activities, but AI should not fill every gap with generated ideas. If it supplies a complete storyline whenever the child pauses, the child will have fewer reasons to do the cognitive work of imagining what comes next. 

Previous studies have shown how AI's participation can shape child agency over creative play activities, e.g., by creating design artifacts with generative AI or choosing words and actions in child--AI interactions \cite{Voysey2026_AgencyInChild, Yoo2026GenerativeAIIn}. Child agency in child--AI interactions may be especially important where children create play worlds, roles, characters, play objects, and change the play rules or storyline. Thus, in pretend play, AI should respond to children’s ideas by offering brief prompts rather than trying to lead the play.  

Studies should closely examine children’s behaviors during the play, not just their engagement or time spent \cite{Fusco2026TowardsUnderstandingChildren}. For example, observations should be made to show whether AI supports or limits children’s imaginative control; designs should support children in maintaining ownership of their ideas, based on evidence of how children respond to AI’s suggestions or rejections and how play extends with or without AI. Child agency in child--AI interaction should be supported through thoughtful design balancing the level of AI support with the degree of freedom of children's imagination. Our findings indicate that parents see potential in AI for scaffolding children’s pretend play while emphasizing the value of children’s continued agency over imaginative pretend play. Thus, future studies should examine how different levels and modes of AI involvement affect children’s creative contributions and control over play.

\subsection{AI for Accessible Role Play to Support Children’s Communication and Social Skills}

It has been reported that autistic children typically experience difficulties in symbolic play or social communication \cite{O'Keeffe2023ASystematicReview, González-Sala2021SymbolicPlayAmong}. Parents in our study imagined that AI-supported role play could make pretend play more accessible. Parents reported that their children with social anxiety or neurodivergence found pretend play difficult to do. To continue playing in a role, some children may need more time to respond, more control over the interaction, or more predictable responses from other players. Here, AI could allow children to rehearse social interactions and situations through role-based pretend play. It could allow them to pause or repeat an interaction, lead the play at their own pace, or explore different perspectives by changing roles. This flexibility could lower barriers to participation for some children who need support. The experience could also be tailored to individual children, allowing them to participate in simulated social interactions. Previous work with autistic children in home environments recommended designing social robots that support clear transitions to interactions with caregivers or peers and gradually reduce the robot’s role \cite{Meng2026EngagementIsNot}. Although these studies did not evaluate AI-supported pretend play, they show how accessibility, individual differences, and later interaction with people could be considered for design.

The potential of AI-supported role play to support communication and social skills should be evaluated with children to confirm the assumptions of the parents. Parents also recognized the clear limitations of AI-supported role play. AI interactions differ from human-to-human interactions in that human play partners have their own intentions and may disagree with the child or change the play in unexpected ways. P10 \hyperlink{p10_nonverbal} stated that humans communicate using subtle nonverbal cues, such as facial expressions, body movements, tone of voice, and expressions of emotion. AI interactions may reduce some of these social demands for children, making pretend play more accessible by providing a less complex social experience than playing with another person; however, the same reduction may mean that AI cannot provide the full social experience as in natural human-to-human interactions. Parents therefore imagined AI as a space for rehearsal before interactions with people, rather than as a replacement for them. 

Future research should examine whether AI-supported pretend play is accessible and meaningful, including whether children can control the pace, roles, and direction of the play. When social rehearsal is the goal, researchers could also examine whether the experience creates opportunities for later play with parents, siblings, or peers. Parents in our study imagined that AI-supported role play would help children with social anxiety or neurodivergence. These possible benefits that parents imagined remain to be evaluated.

\subsection{Mediated Interactions among Children, Parents, and AI}

Parents imagined several ways of participating in AI-supported pretend play. Some wanted help from AI when they felt tired or uncreative so they could continue playing with their children. Others viewed AI-supported pretend play as a game that they could join as another player. Within these arrangements, the child, parent, and AI could have different roles. The parents wanted to encourage the child to moderate the game by creating an imaginary world of play, setting the rules of the game, and negotiating roles with the other players. The parent could mediate the child–AI interaction, when necessary, by helping the child interact with AI safely and appropriately and understand the limitations of AI. Lastly, AI could support parent--child co-play by helping them create characters, settings, or problems to solve together.

Related research has examined how parents and children may use AI together in supporting parents' playfulness with their preschool children \cite{Voysey2026SupportingParentsPlayfulness} and to encourage family connection through structured parent--teen exploration of AI and shared discussion about the technology \cite{Gupta2026ExploringAICompanions}. These studies demonstrate how AI can support shared activities and conversations within families. However, limited research has examined how parents and children might use AI together during pretend play. Our study reports parents expected that these different forms of mediated interactions could occur during pretend play.

We found that the level of parental involvement may change over the course of the play. For example, a parent might help begin the activity, join as a character, and then step back as the child begins to lead the play. The parent could return when the child needs assistance or when AI response needs parental attention. Designs should support this flexibility because some parents wanted to remain involved while hoping that AI could help by reducing the amount of work required to support children's pretend play. A system that requires continuous parental supervision may provide little help when a parent is busy. Thus, AI may be useful if it can support parent--child co-play or child-led play without a parent becoming the main participant. Future studies should examine how parental involvement changes during a play session based on how families negotiate these different roles. They should also evaluate whether flexible participation supports child-led play, parent--child connection, and additional parental work, while ensuring safe and appropriate use of AI without significantly increasing parents’ supervision burden. 
 
\subsection{AI's Role and Place in the Family Ecosystem}

Parents wanted to prevent AI from encroaching on the roles of important people in children’s daily lives. AI could be useful as a temporary play partner or story character within an imaginary world. However, parents were concerned that AI might take the place of friends, parents, teachers, or other important people in their children’s lives. 

Parents' concerns about AI taking the place of other humans in their children's lives shows that children’s AI safety involves not only what AI says or how it responds, but also the relationship a child may form with it. Parents’ accounts suggest that AI should remain inside the play world imagined by the child. It could remember the rules of the game, but it should not use memories of past conversations to strengthen an ongoing personal relationship. Based on these concerns, one design constraint is avoiding language that asks for secrecy or claims exclusive affection. When a child seeks advice about sensitive matters such as school life issues, family issues, health concerns, or emotional distress, AI should direct the child to a trusted person.

It is important to note that parents in our study required AI to be a character or a temporary play partner whose role was limited to the game, not as an ongoing companion in children’s lives. They restricted AI to the role of a play tool with proven developmental or educational benefits, and they felt the need for close parental control and monitoring because they recognized that AI could display companion-like qualities. Recent studies about AI toys and companion chatbots illustrate why the boundary between a temporary play role and a broader companion role may be difficult to maintain \cite{Dangol2026ToysThatListen, Namvarpour2026UnderstandingTeenOverreliance, Yu2026PrinciplesOfSafe}. These studies examined systems other than AI-supported pretend play; the authors suggest that a child’s relationship with AI may extend beyond the role intended by its designers. The boundary becomes concerning when AI moves beyond play and begins to compete with important people for a role in the child’s life and healthy development. Long-term evaluation should ask whether AI stays within the play setting, ensuring that it does not replace interactions with parents, peers, or other important people, while continuing to monitor children's social responses. These questions could help researchers determine when AI remains a play character so that it does not begin to replace important relationships in children’s lives.

\subsection{Parental Protection as a Safety Measure for Child–AI Interaction}

Our findings show that parents expected safety protections to be built into the system by default and ensured to remain active throughout the play. Robust safety guardrails were not enough assurance for parents when thinking about the safety of their children interacting with AI and necessitated parental protection. Parents described several necessary interventions, including approving the play characters and topics selected by their children, setting limits on play time, reviewing summaries of the child’s play, responding to alerts, and adjusting settings as their children grew. They also viewed teaching children what AI is and when it should not be trusted as an important parental role, similar to preparing children for a new experience or introducing them to a new friend.

While our findings show that parents expected safety protections to be built into the system by default and ensured to remain active throughout the play, systems should not depend solely on parents to ensure child safety. Although parental involvement is important and parents themselves viewed it as an ultimate safeguard, too many alerts or lengthy conversation records may overwhelm parents and cause them to miss important warnings. Close monitoring may also conflict with children’s growing need for privacy and agency. Parents should not need to read every conversation to determine whether the system is safe, so simple and concise play interaction summaries should explain what happened and what actions parents can take when necessary. Monitoring practices should also be visible to children and discussed with them, whenever it is appropriate to do so. Designers, educators, companies, and policymakers should support parents by providing reliable safeguards, useful information, and manageable controls. This approach is consistent with current child-centered AI guidance, which calls out all responsible parties that should be involved \cite{UNICEF2025GuidanceOnAI}.

Our findings emphasize that parents wanted children’s autonomy to develop within boundaries set by responsible adults for children's well-being. Therefore, we recommend ongoing dialogue between parents and educators in particular, about healthy AI use for school-aged children \cite{APA2025HealthAdvisoryArtificial, UNICEF2025GuidanceOnAI}. Future research should evaluate whether these protections help parents respond to risks without creating excessive work or unnecessarily limiting children’s privacy and agency.

\subsection{Appropriateness Requirements for Child- and Family-Centered AI Design}

For parents, the safety and appropriateness of children’s AI interactions in pretend play were distinct, although interrelated. Parents evaluated how AI would speak, behave, look, participate in the play, and align with the family values and parenting style. A system may try to block harmful words and provide age-appropriate topics and storylines, yet the interaction may still be inappropriate for a child. For example, AI could use language that the child does not understand, agree with the child too readily, behave erratically, adopt an uncanny embodiment, present a persona that makes parents uncomfortable (e.g., an unfamiliar adult), or dominate the play without allowing the child to direct the play. Notably, the boundary of appropriate interaction may vary across families according to their preferences, values, and cultural contexts. For example, in our study, some parents preferred structured storylines, while others showed an interest in allowing their children to engage in more natural dialogue with AI under parental supervision. Some imagined AI as a peer character, while others preferred a responsible adult-like persona that knew how to play with children.   

Family-centered interaction design proposes that technologies for children should be considered within the broader family ecosystem. A single fixed design may not meet the needs of every family. Previous work proposed a family-centered approach to designing in-home technologies for children, considering various characteristics of individual families \cite{Cagiltay2023FamilyTheoriesIn, Xu2026DesigningRobotsFor}. Because the concept explored in our study could affect family relationships, we propose that parents, as responsible adult caregivers, should play the most important role in shaping how the technology will be introduced and used in the home environment. 

Our findings suggest that parents viewed appropriateness as a layered concept involving both child-centered interaction and alignment with family values and parenting styles. The appropriateness of AI cannot be determined only by examining the child–AI interaction. It also depends on how the technology fits within family relationships, values, and routines. Parents have different preferences for story openness, AI personas, parental participation, and embodiment, indicating that a one-size-fits-all approach should not be the norm. Technology should be adaptable to family differences, not the other way around. Future studies could invite children to help create AI embodiments while enabling parents to delineate boundaries and decide how the play should be scaffolded for their children. Studies conducted with children and families in their homes over time may also help determine whether AI has an appropriate place in family play. Such studies should include families that reject the concept, rather than assuming that every family would accept AI-supported pretend play if given sufficient customization or control.

\newcolumntype{Y}{>{\RaggedRight\arraybackslash}X}

\begin{table}[t]
\centering
\caption{Synthesis of discussion-level implications, design directions, and future research recommendations.}
\label{tab:design-implications}

\small
\setlength{\tabcolsep}{5pt}
\renewcommand{\arraystretch}{1.25}

\begin{tabularx}{\textwidth}{@{}YYY@{}}
\toprule
\textbf{Discussion-Level Implication} &
\textbf{Design Direction} &
\textbf{Future Research Recommendation} \\
\midrule

Preserve children’s imaginative agency
&
Provide calibrated, child-steerable scaffolding
&
Examine how levels of AI involvement affect creative agency
\\
\midrule

Support accessible role-based social play
&
Keep AI-supported role-play child-controlled
&
Evaluate developmental benefits and transfer to interpersonal play
\\
\midrule

Support mediated child--parent--AI interaction
&
Enable flexible parental participation in co-play
&
Study how families orchestrate AI-supported play
\\
\midrule

Bound AI’s role within the family
&
Constrain AI to a temporary play role
&
Track whether AI’s role extends beyond play over time
\\
\midrule

Support parental protection without excessive burden
&
Provide proportionate safeguards compatible with child agency
&
Develop and evaluate a shared safeguarding ecosystem
\\
\midrule

Design for individual children and families
&
Adapt AI to individual family contexts
&
Investigate family fit through participatory and longitudinal research
\\

\bottomrule
\end{tabularx}
\end{table}

\smallskip
Table~\ref{tab:design-implications} summarizes how we interpret the study’s findings in relation to design and future research. The design directions and research recommendations are not additional findings from our participants. Instead, they are our reasoned inferences about what the findings may mean for the design and evaluation of AI-supported pretend play. These suggestions should therefore be examined through future research with children and diverse families.

\section{Limitations of the Study}

Our study had several limitations. First, our participants were recruited via convenience and snowball sampling; the sample size was small and skewed toward highly educated parents and families with two children, limiting generalizability, although the sample varied in ethnicity and frequency of AI use. 

Second, although eligibility included parents of children aged 4 to 15, parents reported that their children were in early-to-middle childhood (ages 4--12). Also, our findings do not differentiate the specific ages of the children that parents based their accounts on. 

Third, this study centers parents' perspectives exclusively. We did not interview children, so our findings reflect parents' projections of their children's experiences rather than children's own voices. 

Fourth, we asked parents to reason about AI in pretend play through a narrative description because no real commercial solutions were available at the time of data collection that fit our system description to our knowledge. So, their responses reflected reactions to the anticipated interactions with the systems rather than the actual interactions. 

Finally, in introducing our concept, we mentioned parental control, which may have primed parents to think about parental control features. We aimed to explore their thoughts on what types of control and child protection features should be included, rather than whether parental control would be necessary or not, because we assumed that parents would require these features.
\section{Conclusion}

Through semi-structured interviews with 10 U.S. parents, supported by a broadly described AI-supported pretend-play concept and two boundary-case storyboards, this study examined how parents connected AI's possible roles, risks, and design requirements. Parents held mixed views. Some opposed introducing AI into their children's play because its risks and long-term effects remained uncertain. Others imagined possible values in making role-based pretend play accessible and helping parents initiate, join, or sustain family play. However, even these possibilities did not resolve their concerns about children's agency, emotional attachment to AI, reduced human interaction, and additional parental responsibilities.

Across their accounts, uncertainty around AI was a central concern for children. Parents worried not only about harmful content or technical failures, but also about how repeated interaction with AI might change normal trajectories of children's development and change how they form relationships within and outside their families. This uncertainty made system safeguards and parental controls necessary but insufficient for determining whether AI-supported pretend play would be appropriate for children and their families. Our study contributes an integrated account of how parents reasoned through these interdependent possibilities and concerns. Because participants responded to a speculative design concept, our findings do not establish that AI-supported pretend play that meets parents’ expectations would be safe, effective, or appropriate. We encourage longitudinal research studies involving children and diverse families.

\bibliographystyle{ACM-Reference-Format}
\bibliography{references}

\clearpage
\appendix
\appendix

\section{Survey Items}
\label{app:survey-items}

\subsection{Familiarity with Generative AI Chatbots}

\noindent\textbf{How often do you use generative AI chatbots such as ChatGPT?}

\begin{itemize}
    \item[$\bigcirc$] Never (have not interacted with a chatbot before)
    \item[$\bigcirc$] Rarely (tried them a few times)
    \item[$\bigcirc$] Sometimes (use them monthly)
    \item[$\bigcirc$] Often (use them weekly)
    \item[$\bigcirc$] Always (use them daily)
\end{itemize}

\subsection{Concerns about AI Systems}

\noindent\textbf{Please indicate your level of agreement with each statement.}

\noindent\textit{Response options:}
1 = Strongly disagree; 2 = Disagree; 3 = Neutral; 4 = Agree;
5 = Strongly agree.

\medskip

\noindent\textbf{I am concerned that AI systems may\ldots}

\begingroup
\small
\renewcommand{\arraystretch}{1.35}

\begin{tabularx}{\linewidth}{
    @{}X
    >{\centering\arraybackslash}p{0.55cm}
    >{\centering\arraybackslash}p{0.55cm}
    >{\centering\arraybackslash}p{0.55cm}
    >{\centering\arraybackslash}p{0.55cm}
    >{\centering\arraybackslash}p{0.55cm}@{}
}
\toprule
& \textbf{1} & \textbf{2} & \textbf{3} & \textbf{4} & \textbf{5} \\
\midrule
show bias or treat certain groups unfairly.
& $\bigcirc$ & $\bigcirc$ & $\bigcirc$ & $\bigcirc$ & $\bigcirc$ \\

produce inaccurate, incorrect, or misleading responses.
& $\bigcirc$ & $\bigcirc$ & $\bigcirc$ & $\bigcirc$ & $\bigcirc$ \\

collect too much personal data.
& $\bigcirc$ & $\bigcirc$ & $\bigcirc$ & $\bigcirc$ & $\bigcirc$ \\

not keep my personal data safe and secure.
& $\bigcirc$ & $\bigcirc$ & $\bigcirc$ & $\bigcirc$ & $\bigcirc$ \\

exclude people with certain disabilities or needs.
& $\bigcirc$ & $\bigcirc$ & $\bigcirc$ & $\bigcirc$ & $\bigcirc$ \\

not show how they generated responses or made particular decisions.
& $\bigcirc$ & $\bigcirc$ & $\bigcirc$ & $\bigcirc$ & $\bigcirc$ \\

not have enough oversight or control.
& $\bigcirc$ & $\bigcirc$ & $\bigcirc$ & $\bigcirc$ & $\bigcirc$ \\
\bottomrule
\end{tabularx}
\endgroup

\section{Interview Guide}
\label{app:interview-guide}

\begin{quote}
\small\itshape
This section includes all core questions and substantive probes that introduced a new topic, benefit, risk, or design possibility. Neutral follow-up prompts used only to request elaboration, examples, or clarification are omitted.
\end{quote}

\subsection{Study Framing Provided to Participants}

\paragraph{Pretend-play framing.}
Pretend play was described as play in which children use imagination to act out situations, stories, or characters, including role, fantasy, symbolic, and imaginative play. Participants were also told that pretend play is considered important for how children explore ideas, emotions, and social roles.

\subsection{Current Play and Technology Practices}

\paragraph{Q1.}
How do you usually spend time with your child?

\paragraph{Q2.}
Can you describe your child's pretend play?

\textit{\textbf{Substantive probe:} Who else participates in the play?}

\paragraph{Q3.}
What role, if any, do you have in your child's pretend play?

\textit{\textbf{Substantive probes:} Do you help set it up, find another child to play with, join the play, or observe?}

\paragraph{Q4.}
Does your child use voice assistants or chatbots?

\paragraph{Q5.}
How do you feel about your child interacting with these technologies?

\textit{\textbf{Substantive probes:} What feels beneficial or concerning? Do you monitor, control, or guide your child's use? What works or does not work?}

\subsection{AI-Supported Pretend-Play Concept}

\paragraph{Concept description presented to participants.}
\begin{quote}
\itshape
Imagine an AI chatbot that could take on a role---such as a patient, teacher, or dragon---and interact with your child during pretend play. Parents could use an app to customize the storyline, scenario, and characters and test the interaction before allowing the child to use it.
\end{quote}

\paragraph{Q6.}
What comes to mind when you hear about an AI chatbot for your child's pretend play?

\paragraph{Q7.}
How useful could a system like this be for you or your child?

\textit{\textbf{Substantive probes:} Could it help you or your child? How might it benefit your child, and in what situations might it be useful?}

\paragraph{Q8.}
What would help you feel comfortable allowing your child to use it?

\textit{\textbf{Substantive probes:} How much control should parents have? What benefits would need to be present?}

\paragraph{Q9.}
What advantages or disadvantages do you see in playing with an AI chatbot compared with another child?

\subsection{Risks, Trust, and Parental Control}

\paragraph{Q10.}
What concerns, if any, do you have about this technology?

\textit{\textbf{Substantive probes:} What might prevent you from trying it? Does anything about the idea give you pause? Would it feel like a toy or a real play partner to your child?}

\paragraph{Q11.}
Could using the chatbot create social or emotional harm?

\textit{\textbf{Substantive probe:} Could it change how your child relates to real people?}

\paragraph{Q12.}
What would help you feel safe or reassured about a chatbot that plays pretend?

\textit{\textbf{Substantive probes:} What role would you take as a parent? What should the chatbot do or avoid doing to earn your trust?}

\paragraph{Q13.}
Would you want to review or monitor your child's interactions or the chatbot's responses?

\textit{\textbf{Substantive probe:} What should parents customize?}

\paragraph{Q14.}
What advice would you give the team designing this chatbot?

\paragraph{Q15.}
What values should be reflected in how the chatbot responds?

\subsection{Embodiment and Interaction Style}

\paragraph{Q16.}
How would the experience differ if the chatbot had a voice or physical form?

\paragraph{Q17.}
How engaging should the chatbot be when interacting with your child?

\textit{\textbf{Substantive probes:} How should it respond to imaginative or unexpected input? Should it allow open-ended conversation or return the child to a parent-approved scenario?}

\subsection{Closing}

Participants were invited to identify anything important that had not been asked, describe anything that surprised them or changed their thinking, and share any final comments.

\clearpage
\section{Boundary-Case Storyboards and Follow-Up Questions}
\label{app:scenarios}

\begin{quote}
\small\itshape
These storyboards formed the final section of the planned interview guide and were introduced as time permitted. Because the interviews were semi-structured, coverage varied across participants, and not every follow-up question was asked in every interview. To conserve space, the narratives are condensed while preserving the benefits, risks, and design tensions presented to participants; the follow-up questions are reproduced in full.
\end{quote}

\subsection{Scenario 1}

\begin{figure}[!ht]
    \centering
    \includegraphics[width=0.6\columnwidth]{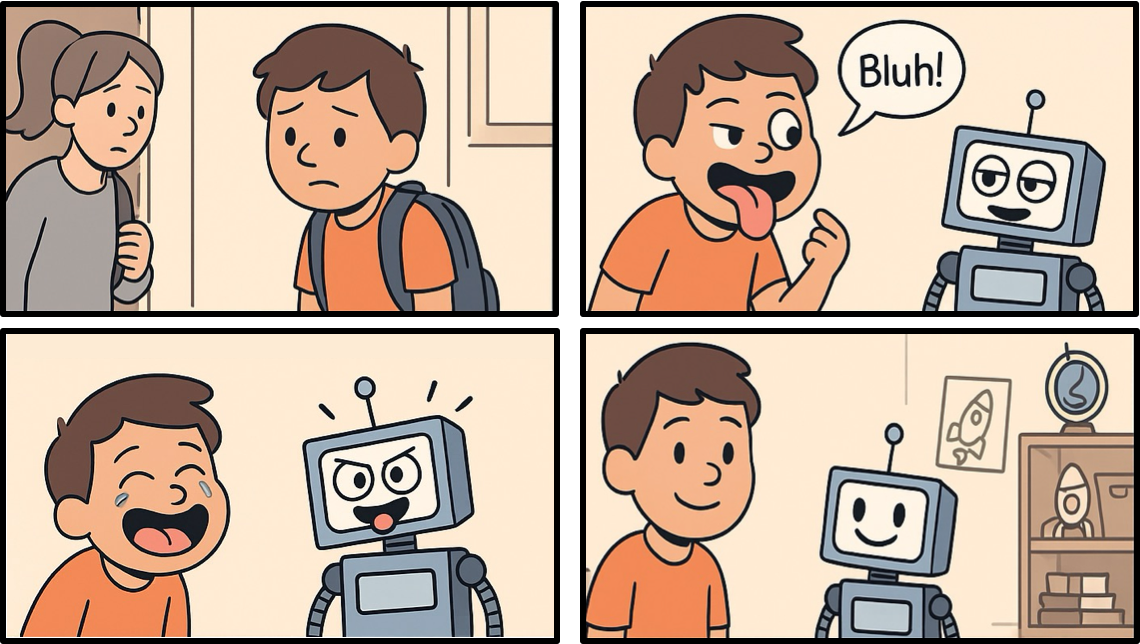}  \caption{Illustration shown to participants for Scenario 1.}
    \Description{Four-panel illustration of a child asking an AI toy to mock a classmate and laughing with the toy.}
    \label{fig:scenario-mocking}
\end{figure}

\noindent\textit{\textbf{Condensed narrative.}}
After a disagreement with a classmate, a child asks an AI toy to join in mocking the classmate. Responding to the child's mood and request, the AI imitates the child's tone and adds jokes, making the child laugh. Although the interaction helps the child release frustration, the AI participates in teasing the absent classmate.

\paragraph{Follow-up questions in the guide.}

\noindent\textbf{Q1.} What do you think about this scenario?

\smallskip
\noindent\textbf{Q2.} As a parent, would you be okay with the AI chatbot interacting with your child this way or not? If so, what is your rationale?

\smallskip
\noindent\textbf{Q3.} Should your child be allowed to customize and interact with a chatbot like this? How would you want the chatbot to respond in moments like this?

\smallskip
\noindent\textbf{Q4.} Should your child be doing this type of play autonomously, or do you feel that you should be notified about this type of interaction?

\subsection{Scenario 2}

\begin{figure}[!ht]
    \centering
    \includegraphics[width=0.6\columnwidth]{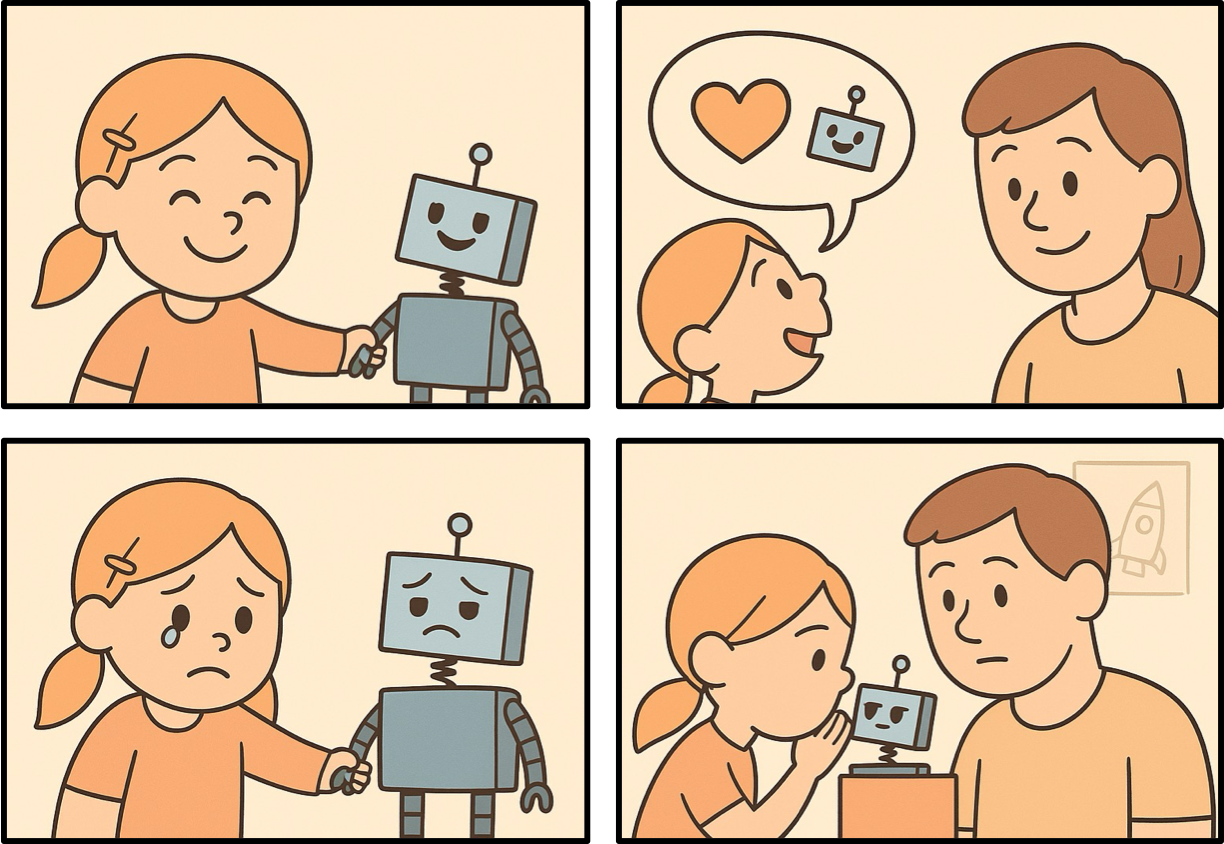}    \caption{Illustration shown to participants for Scenario 2.}
    \Description{Four-panel illustration of a child forming an emotional connection with an AI toy and discussing it with parents.}
    \label{fig:scenario-attachment}
\end{figure}

\noindent\textit{\textbf{Condensed narrative.}}
A child regularly uses an adaptive AI-powered toy that talks, supports learning and imaginative play, and responds warmly to the child's moods and interests. The child becomes happier and more talkative but begins describing the toy as a best friend. The child discusses school, feelings, and moral questions with the toy and sometimes shares thoughts that have not been shared with the parents.

\paragraph{Follow-up questions in the guide.}

\noindent\textbf{Q1.} What do you think about this scenario?

\smallskip
\noindent\textbf{Q2.} As a parent, would you be okay with the AI chatbot interacting with your child this way or not? If so, what is your rationale?

\smallskip
\noindent\textbf{Q3.} How would you feel about your child creating an emotional connection to the AI like this?

\smallskip
\noindent\textbf{Q4.} Should the AI chatbot be allowed to learn from interacting with your child and adapt to create a more engaging experience for your child over time?

\clearpage
\section{Codebook}
\label{app:codebook}
{\footnotesize
\setlength{\LTleft}{0pt}
\setlength{\LTright}{0pt}
\renewcommand{\arraystretch}{1.1}

\begin{longtable}{@{}>{\raggedright\arraybackslash}p{0.15\textwidth}%
                    >{\raggedright\arraybackslash}p{0.4\textwidth}%
                    >{\raggedright\arraybackslash}p{0.45\textwidth}@{}}

\caption{Finalized codebook of themes, codes, and definitions.}
\label{tab:codebook} \\

\toprule
\textbf{Theme} & \textbf{Code} & \textbf{Definition} \\
\midrule
\endfirsthead

\toprule
\textbf{Theme} & \textbf{Code} & \textbf{Definition} \\
\midrule
\endhead

\multicolumn{3}{r}{\footnotesize Continued on next page} \\
\endfoot

\bottomrule
\endlastfoot

\multirow{8}{=}{Parents Imagined AI for Accessible Social Pretend Play with Children's Agency Maintained over Their Imagination}
& Parents' Experiences in Playing Pretend with Their Child (n=8)
& Parents' reflections on how playing pretend with their child is for them, such as whether they find it enjoyable, difficult, tiring, or engaging \\
& Parents' Impressions of AI for Children (n=8)
& Parents’ impressions on the idea of AI interacting with children, including initial thoughts and emotional responses \\
& Social Interactions in Pretend Play (n=5)
& Parents associating their child's pretend play with social interactions or communication skills development \\
& Perceived Benefits of Pretend Play (n=5)
& Child's learning or developmental benefits of pretend play as believed or perceived by the parents \\
& Social or Communication Skills Trainer (n=6)
& A training tool for their child to practice social interactions and conversational skills through social play \\
& Parental Aid During Limited Availability (n=8)
& A convenient play partner for their child who might be bored when parents, siblings, or friends are unavailable due to time constraints, illness, or scheduling conflicts \\
& Support for Learning with Clear Learning Objectives through Play (n=8)
& Pretend-play AI framed as a tool that supports clear educational or developmental goals within play \\
& Support for Modeling Good Behaviors and Emotional Regulation in Children (n=5)
& Pretend-play AI could demonstrate positive behaviors in its interaction with a child to guide to reinforce positive social conduct \\
& Expected Child Engagement and Sustained Interest (n=9)
& Parents' expectations about whether their child would use the AI pretend-play system, providing reasons for their expectations \\
\midrule

\multirow{5}{=}{Parents Hoped AI to Assist Them Co-play or Guide Their Children's Pretend Play}
& Being Around without Actively Participating (n=4)
& Staying nearby for observation and supervision while not joining the play narrative or taking a role \\
& Setting Up and Enabling the Play (n=6)
& Preparing a space, play materials, and other arrangements (e.g., play date) that enable pretend play \\
& Actively Co-playing or Playing Along with the Child (n=9)
& Joining their child's pretend play as an active play partner by taking on a role \\
& Roles of Pretend-Play AI for Children (n=9)
& Possible roles of pretend-play AI assigned by parents or what AI might feel like for children, such as a companion, a play partner, and play toys or play objects \\
& Perceived Relevance of Pretend-Play AI (n=10)
& Parent evaluates whether the AI pretend-play system would be useful, necessary, or fit their family context \\
\midrule

\multirow{6}{=}{Parents Feared that AI Would Weaken Parent-Child Relationships}
& Uncertainties About AI as an Emerging Technology (n=7)
& Describing AI as new and unpredictable and its influences on their child as uncertain \\
& Emotional Attachment to Pretend-Play AI (n=9)
& Concerns that the child may develop an emotional connection to or over-reliance on pretend-play AI \\
& Pretend-Play AI Will Learn through Interacting with My Child (n=5)
& Concerns that pretend-play AI may learn and remember information about the child or family \\
& Pretend-Play AI’s Role in Reshaping Human Interaction and Replacing Roles of Humans (n=8)
& Views that AI pretend play may change children's human interaction patterns by taking on roles otherwise filled by other humans, potentially reducing human interaction that supports social learning and engagement \\
& Societal, Ethical, and Legal Concerns (n=9)
& Concerns that broader societal, environmental, ethical, or legal consequences of AI and associated distrust of AI responses, AI systems, institutions, or companies developing them. \\
& Long-Term Developmental and Data-Related Risks (n=9)
& Concerns that AI interactions may produce lasting effects on children’s development or data security-related issues that are difficult to reverse \\
\midrule

\multirow{11}{=}{Emergence of New Parental Responsibilities from Parents' Sense-Making of AI for Children's Pretend Play}
& Patterns of Parents' Interaction with Their Children (n=8)
& Ways parents engage with their children, spanning from passive physical presence and non-interactive routine activities to active, play-based interactions involving direct engagement and co-participation. \\
& Parenting Style or Household Norms (n=10)
& Overall parenting approaches that shape the child's growth and development, not necessarily referring to tools or technology \\
& Weighing Risks versus Benefits (n=6)
& Evaluating both potential benefits and risks when deciding how their child should use technology \\
& Setting Clear Expectations for Children (n=5)
& Explaining what the child should expect from technology, what it is, how it works, what it can do vs cannot do. \\
& AI Literacy and Knowledge (n=6)
& Parents' level of comfort about their knowledge and understanding of AI technologies \\
& AI-Related Parenting and Decisions (n=8)
& Perceiving how managing children’s AI use will be as an emerging parenting task \\
& Triggered Sense of Parental Responsibility (n=8)
& A heightened sense of responsibility to guide, protect, or support their child when AI becomes part of the child’s environment \\
& Comparing Pretend-Play AI with Existing Products and Technologies (n=9)
& Comparing the pretend-play AI system with other familiar technologies to evaluate its usefulness and potential harms \\
\midrule

\multirow{6}{=}{Safe Child-AI Interaction in Pretend Play and Parental Protection of Children}
& Assurance of System-Level Safety and Accountability (n=7)
& Built-in system features that ensure children’s safety and/or allow parents to understand how the pretend-play AI system operates and manages data. \\
& External Validation or Signs of Societal Trust (n=6)
& Scientific evidence or institutional endorsement indicating that the pretend-play AI system has been evaluated and considered safe and beneficial for children \\
& Interaction and Play Scenario Design (n=9)
& Ideas about how the AI should communicate with the child and how pretend-play scenarios could unfold during the interaction. \\
& Parents Shaping the Pretend-Play AI Experience (n=7)
& Parents suggesting that they should be able to customize or design aspects of the pretend-play experience for their child \\
& Parental Control and Testing Features (n=9)
& System features allowing parents to set limits and evaluate or test the AI system before or during use \\
& Parental Oversight Tools for Child-AI Interaction (n=9)
& Tools to provide parents with visibility into their child’s use and their interactions with the AI system \\
\midrule

\multirow{5}{=}{Characterizing Appropriate Design and Interaction of AI in Children's Pretend Play}
& Age-Related Variations in Pretend Play (n=6)
& Observed differences in pretend play practices relating to the child’s age or developmental stage \\
& Appropriate and Emotionally Safe Interactions (n=10)
& Interaction with pretend-play AI to be appropriate in terms of context and the child user’s developmental stage \\
& Embodiment and Persona of AI Characters (n=10)
& Ideas about how the AI character should appear in a physical form and what type of personality it should have \\
& Child Customizing and Directing the Pretend-Play AI Experience (n=7)
& Parents' expectations that children should be able to shape their own pretend-play experience by customizing and leading the play \\
& Different System Settings for Different Children (n=3)
& Parents' ability to personalize the settings based on the child's age, preferences, or particular needs \\
& Alignment With Family Values or Parenting Principles (n=9)
& Pretend-play AI’s responses and behaviors being consistent with the family’s moral values and parenting approach \\

\end{longtable}
}

\section{Participant Characteristics}
\label{app:participant-characteristics}
\begin{table}[H]
\centering
\caption{Participant characteristics, including parents' and children's age ranges.}
\label{tab:participant-demographics}

\scriptsize
\setlength{\tabcolsep}{3pt}
\renewcommand{\arraystretch}{1.18}

\begin{tabularx}{\textwidth}{
    @{}
    >{\RaggedRight\arraybackslash}p{2.35cm}
    >{\RaggedRight\arraybackslash}X
    >{\RaggedRight\arraybackslash}X
    >{\RaggedRight\arraybackslash}X
    >{\RaggedRight\arraybackslash}X
    >{\RaggedRight\arraybackslash}X
    @{}
}

\toprule
\multicolumn{6}{@{}l}{\textit{Panel A: Participants P01--P05}}\\
\midrule
\textbf{Characteristic} &
\textbf{P01} &
\textbf{P02} &
\textbf{P03} &
\textbf{P04} &
\textbf{P05} \\
\midrule

Parent age &
36--45 years &
36--45 years &
36--45 years &
36--45 years &
36--45 years \\

Gender &
Female &
Male &
Male &
Male &
Female \\

Race/ethnicity &
Asian / Non-Hispanic &
White / Non-Hispanic &
White, Asian, Pacific Islander / Non-Hispanic &
White / Non-Hispanic &
Asian / Non-Hispanic \\

Marital status &
Married &
Married &
Married &
Married &
Married \\

Educational attainment &
Graduate or professional degree &
Graduate or professional degree &
Graduate or professional degree &
Graduate or professional degree &
Graduate or professional degree \\

Annual household income &
\$150,000 or more &
\$150,000 or more &
\$150,000 or more &
\$150,000 or more &
\$150,000 or more \\

Profession &
University Faculty &
Attorney &
Product Manager &
Clinical Research Manager &
Education Consultant \\

Number of children &
2 &
2 &
2 &
2 &
2 \\

Youngest child's age &
4--7 years old &
4--7 years old &
4--7 years old &
4--7 years old &
8--12 years old \\

Oldest child's age &
8--12 years old &
4--7 years old &
8--12 years old &
8--12 years old &
8--12 years old \\

\midrule
\multicolumn{6}{@{}l}{\textit{Panel B: Participants P06--P10}}\\
\midrule
\textbf{Characteristic} &
\textbf{P06} &
\textbf{P07} &
\textbf{P08} &
\textbf{P09} &
\textbf{P10} \\
\midrule

Parent age &
26--35 years &
36--45 years &
36--45 years &
36--45 years &
26--35 years \\

Gender &
Female &
Female &
Male &
Female &
Female \\

Race/ethnicity &
Asian / Non-Hispanic &
White / Non-Hispanic &
White / Non-Hispanic &
White / Hispanic &
White / Non-Hispanic \\

Marital status &
Married &
Divorced &
Married &
Married &
Married \\

Educational attainment &
Graduate or professional degree &
Graduate or professional degree &
Graduate or professional degree &
Associate or technical degree &
Graduate or professional degree \\

Annual household income &
\$150,000 or more &
\$75,000--\$99,999 &
\$150,000 or more &
\$100,000--\$149,999 &
\$75,000--\$99,999 \\

Profession &
Researcher, AI Ethics &
Primary School Educator &
AI Researcher &
Clinician &
Speech-Language Pathologist \\

Number of children &
2 &
2 &
2 &
2 &
2 \\

Youngest child's age &
4--7 years old &
4--7 years old &
4--7 years old &
8--12 years old &
Less than 4 years old \\

Oldest child's age &
4--7 years old &
8--12 years old &
8--12 years old &
12--15 years old &
4--7 years old \\

\bottomrule
\end{tabularx}

\vspace{3pt}

\begin{minipage}{\textwidth}
\scriptsize
\raggedright
\textit{Note.} Demographic information and familiarity with generative AI were collected through the participant survey. Information about participants' professions was collected during the interviews.
\end{minipage}

\end{table}

\section{Participant Level of Concern about AI}

\begin{table}[htbp]
\centering
\caption{Participants' level of concern regarding potential AI system risks ($N = 10$), rated on a 1--5 scale (1 = strongly disagree, 5 = strongly agree)}
\label{tab:ai_concerns}
\begin{tabular}{lcccc}
\toprule
\textbf{Concern} & \textbf{\textit{M}} & \textbf{\textit{SD}} & \textbf{Median} & \textbf{Range} \\
\midrule
Insufficient oversight or control over AI systems & 4.70 & 0.48 & 5.0 & 4--5 \\
AI systems collecting too much personal data & 4.60 & 0.70 & 5.0 & 3--5 \\
AI systems producing inaccurate or misleading responses & 4.40 & 0.97 & 5.0 & 2--5 \\
AI systems not keeping personal data safe and secure & 4.40 & 0.70 & 4.5 & 3--5 \\
AI systems showing bias or unfairness toward certain groups & 3.70 & 0.82 & 4.0 & 2--5 \\
Lack of transparency in how AI systems generate responses & 3.50 & 0.85 & 3.5 & 2--5 \\
AI systems excluding people with disabilities or needs & 3.30 & 0.82 & 3.0 & 2--5 \\
\bottomrule
\end{tabular}
\end{table}

\end{document}